\documentclass[draft,tightenlines,nofootinbib,preprint,aps,eqsecnum,amsmath,amssymb]{revtex4}

\newcommand{\beq}{\begin{equation}}
\newcommand{\eeq}{\end{equation}}
\newcommand{\bea}{\begin{eqnarray}}
\newcommand{\eea}{\end{eqnarray}}
\newcommand{\cir}{{\buildrel \circ \over =}}
\newcommand{\sgn}{\epsilon}
\newcommand{\eo}{{}^4{\buildrel \circ \over E}}

\begin{document}

\title{Hamiltonian Expression of Curvature Tensors in the York Canonical Basis:
I) The Riemann Tensor and Ricci Scalars}

\medskip

\author{Luca Lusanna}

\affiliation{ Sezione INFN di Firenze\\ Polo Scientifico, Via Sansone 1,
50019 Sesto Fiorentino (FI), Italy\\ E-mail: lusanna@fi.infn.it}

\author{Mattia Villani}

\affiliation{Dipartimento di Fisica, Universita' di Firenze\\ Polo Scientifico, Via Sansone 1,
50019 Sesto Fiorentino (FI), Italy\\ E-mail: villani@fi.infn.it}

\today

\bigskip

\begin{abstract}

By using the York canonical basis of ADM tetrad gravity, in a formulation using radar 4-coordinates for the parametrization of the 3+1 splitting of the space-time, it is possible to write the 4-Riemann
tensor of a globally hyperbolic, asymptotically Minkowskian space-time as a Hamiltonian tensor, whose components are 4-scalars with respect to the ordinary world 4-coordinates,
plus terms vanishing due to Einstein's equations. Therefore "on-shell" we find the expression of the
Hamiltonian 4-Riemann tensor. Moreover, the 3+1 splitting of the space-time used to define the phase space allows us to introduce a Hamiltonian set of null tetrads and to find the Hamiltonian expression of the 4-Ricci scalars of the Newman-Penrose formalism.

This material will be used in the second paper to study the 4-Weyl tensor, the 4-Weyl scalars and the four Weyl eigenvalues and to clarify the notions of Dirac and Bergmann observables.

\end{abstract}

\maketitle

\vfill\eject

\section{Introduction.}

In general relativity (GR) the curvature tensors of the space-time solution of
Einstein equations (or of their existing modifications) play a basic role connecting the
structure of the space-time to the matter present in it. The geometrical and tensorial properties of the Riemann, Ricci and Weyl tensors are well understood and are exposed in all the books on GR. These tensors are functions
of the 4-metric tensor (and of its gradients) of the space-time, which is the geometrical quantity
describing the gravitational field. However the Hamiltonian expression of these tensors has still to be investigated.
\medskip

At the Lagrangian level we have the Hilbert action implying Einstein's equations as its Euler-Lagrange equations. This action is invariant under the local Noether transformations generated by the group of passive diffeomorphisms (world 4-coordinate transformations) of the space-time 4-manifold. These Lagrangian gauge transformations imply that only two combinations of Einstein's equations depend on the second time derivatives of the 4-metric (the accelerations). As a consequence the general covariance of the theory implies that 8 of the 10 degrees of freedom described by the 4-metric have to be considered as Lagrangian gauge variables. However we are still lacking the identification at the Lagrangian level not only of these 8 gauge variables but also of the two physical (i.e. invariant under Lagrangian gauge transformations) degrees of freedom of the gravitational field as 4-scalar functionals of the 4-metric. Only in the linearized theory in harmonic gauges they are known  to be connected with the two polarizations of the
gravitational waves.
\medskip

Only at the Hamiltonian level in the canonical ADM formulation of GR \cite{1} there is a technology (Dirac's
theory of constraints \cite{2} and the Shanmugadhasan canonical transformations \cite{3}) for facing the problem of the identification of the two pairs of canonical variables describing the  independent physical degrees of freedom of the gravitational field (the so-called Dirac observables (DO)). These two pairs of conjugated DO's are functionals of the ten fields of the 4-metric and of the ten conjugate momenta invariant under the Hamiltonian gauge transformations generated by the 8 (four primary and four secondary) first class constraints of the theory. Since the definition of phase space requires a 3+1 splitting of the space-time (an identification of what is "time" and what is "3-space" inside the Lorentzian 4-manifold), the Hamiltonian gauge transformations act on the so foliated space-time and their Legendre pull-back is not the Lagrangian group of the passive 4-diffeomorphisms of the space-time, but rather a group of dynamical symmetries of Einstein's equations having a non-null intersection with the Lagrangian Noether transformations (the 3-diffeomorphisms of the chosen 3-spaces).

Besides the DO's there are eight (four primary and four secondary) Hamiltonian {\it inertial gauge} degrees of freedom with the conjugate variables determined by the eight first class constraints of canonical ADM gravity. These Hamiltonian gauge variables describe the freedom in the choice of the non-inertial (due to the equivalence principle) reference frames centered on a time-like observer, in particular the choice of the instantaneous (generically non Euclidean) 3-spaces labeled by some time variable and parametrized with a set of 3-coordinates. The fixation of a gauge is made
by adding four primary gauge-fixing constraints for the secondary first-class constraints: they determine the primary gauge variables. Then the preservation in time of these gauge fixings generates the secondary gauge fixings for the secondary gauge variables.  At the Lagrangian level the analogue of the gauge fixings for the Hamiltonian gauge variables are the four coordinate conditions (and their preservation in time) selecting a world 4-coordinate system (think to the harmonic gauge conditions). Therefore in a completely fixed gauge the world 4-coordinate system is completely determined both at the Lagrangian and Hamiltonian levels.

\medskip

The non-linearity of the theory is the main obstacle to the completion of the program to find the gravitational DO's \footnote{Only existence theorems have been found for DO's \cite{4}, but not any explicit construction of them.}. In particular it is not yet clear whether the DO's are tensorial quantities, i.e. whether they are 4-scalars like in the old proposal of Bergmann \cite{5} (the so-called Bergmann observables (BO) of the gravitational field; see also Ref.\cite{6,7}).

\bigskip

In Ref.\cite{8} there is a new Hamiltonian formulation of ADM tetrad gravity \footnote{One uses the ADM Lagrangian with the 4-metric decomposed in terms of cotetrads.}, which is needed if the matter contains fermions, for the family of globally hyperbolic, asymptotically flat space-times without super-translations. In them  global non-inertial frames were defined by giving the world-line of a time-like observer and  3+1 splittings of the space-time, namely nice foliations with instantaneous 3-spaces (diffeomorphic to $R^3$), and by using Lorentz-scalar radar 4-coordinates \cite{9,10,11} instead of the standard world 4-coordinates. The transition functions from the world 4-coordinates to the radar one's transform world tensors into radar tensors, whose components are 4-scalars of the space-time.

The phase space contains 16 fields (the components of the cotetrads) and 16 conjugate momenta. There are 14 first class constraints (10 primary and 4 secondary), so that there are 14 Hamiltonian inertial gauge variables and two pairs of Hamiltonian physical degrees of freedom (the tidal variables). In Ref.\cite{12} it was found a Shanmugadhasan canonical transformation to a canonical basis (the York basis \footnote{It is called York basis because one of the inertial gauge variables is the trace of the extrinsic curvature of the 3-spaces as sub-manifolds of the space-time: this quantity is known as the York time \cite{13}.}) adapted to the ten primary first class constraints, but not to the secondary ones (the super-Hamiltonian and super-momentum constraints). In the York basis the two physical degrees of freedom are the two eigenvalues of the spatial 3-metric with determinant one. They are 3-scalars of the 3-space, 4-scalars of the space-time  and are DO's only with respect to the ten primary first class constraints. In a completely fixed Hamiltonian gauge there is a uniquely determined system of radar 4-coordinates and an associated system of world ones.
\medskip

If one would be able to solve the super-Hamiltonian constraint (the Lichnerowicz equation) in its unknown, i.e. the determinant of the spatial 3-metric, and to solve the three super-momentum constraints in their unknowns, i.e. three suitable momenta of the York basis, it would be possible to find a Shanmugadhasan canonical transformation to a final canonical basis adapted to all the 14 first class constraints. The two pairs of physical tidal variables in this basis would be a set of real  DO's for canonical gravity with the property of being also BO's, being 4-scalar due to the use of radar 4-coordinates.

\medskip

In the three papers of Ref.\cite{14,15,16} (quoted as I, II, III, in the sequel) there is the York basis of ADM canonical gravity plus the electro-magnetic field plus positive-energy scalar particles. In the papers II and III there is the weak field Hamiltonian Post-Minkowskian (HPM) limit of the theory and its study in the family of (non-harmonic) 3-orthogonal Schwinger time gauges with applications to gravitational waves and dark matter.

\bigskip

In this paper we will first show that the canonical variables of the York basis, living in the 3-spaces of the 3+1 splitting of the space-time centered on a time-like observer are 4-scalars of the space-time due to the use of radar 4-coordinates.Therefore if these special coordinates are used in place of the ordinary world 4-coordinates we have a framework in which DO's and BO's for the gravitational field coincide.

\medskip

Then we will find the Hamiltonian expression of the 4-Riemann tensor  in terms of the cotetrads (and of their gradients) of ADM tetrad gravity. By using Einstein's equations for expressing the 4-Ricci tensor in terms of the energy-momentum tensor of the matter (its Hamiltonian expression is known if there are not derivative couplings), we can express the 4-Riemann tensor as the sum of a Hamiltonian Riemann radar tensor in the York basis plus terms vanishing by using Einstein's equations (or their equivalent Hamilton equations in phase space). Therefore the Hamiltonian Riemann  radar tensor becomes  equal to the Riemann tensor only {\it on-shell} on the solution of the equations of motion.

\medskip

Moreover, we can define a set of Hamiltonian null tetrads in the York basis and to get the Hamiltonian expression of Ricci scalars. This open the possibility to have a Hamiltonian reformulation of the Newman-Penrose formalism (see for instance Ref.\cite{17}).

\bigskip

The results of this paper will be used in the second paper to get the Hamiltonian Weyl radar tensor (equal to the 4-Weyl tensor on-shell) and the Hamiltonian expression of the Weyl scalars of the Newman-Penrose formalism. This will also allow us to get the Hamiltonian expression of the four Weyl eigenvalues (4-scalars independent from the choice of the null tetrads), which have been used in Ref. \cite{7} to give a physical identification of the mathematical points of the space-time 4-manifold. They will be shown to be neither DO's nor BO's due to their dependence on the Hamiltonian inertial gauge variables.
\bigskip

In Section II we make an extended review of ADM tetrad gravity in the York canonical basis,
with an explicit clarification of the 4-scalar nature  of its variables due to the use of radar 4-coordinates in the 3+1 splitting of the space-time needed to define the phase space. Moreover we express the  4-Ricci tensor as the sum of Einstein's equations plus the Hamiltonian expression of the energy-momentum tensor of the matter.

In Section III we give the Hamiltonian expression of the 4-Christoffel symbols and the we find the Hamiltonian 4-Riemann radar tensor, which is equal to the 4-Riemann tensor plus terms vanishing when we impose Einstein's equations.

In Section IV we introduce a Hamiltonian set of null tetrads, which allows us to get the Hamiltonian expression of the 4-Ricci scalars of the Newman-Penrose formalism.

Then there are some Conclusions and Appendix A containing the Hamiltonian expression of the 3-Riemann and 3-Ricci tensors of the 3-spaces of the 3+1 splitting of the space-time.

\section{ADM Tetrad Gravity in the York Canonical Basis: the Kinematical Background}

In this Section we make a review of the formulation of canonical ADM tetrad gravity
in globally hyperbolic, topologically trivial,
asymptotically Minkowskian space-times without super-translations given in
Refs. \cite{12,14,15,16}. These space-times must also be without Killing symmetries, because,
otherwise, at the Hamiltonian level one should introduce complicated
sets of extra Dirac constraints for each existing Killing vector.
Moreover the spatial 3-metric must have three distinct eigenvalues
to avoid degenerate cases which can eventually be reached by adding the equality of two eigenvalues
as first class constraints by hand. In this class of space-times the ten {\it strong} asymptotic ADM
Poincar\'e generators $P^A_{ADM}$, $J^{AB}_{ADM}$ (they are fluxes
through a 2-surface at spatial infinity) are well defined
functionals of the 4-metric fixed by the boundary conditions at
spatial infinity and of matter (when present). These ten strong generators can be
expressed in terms of the weak asymptotic ADM Poincar\'e generators (integrals on
the 3-space of suitable densities) plus first class constraints.
The absence of super-translations implies that the ADM 4-momentum is asymptotically orthogonal to the
instantaneous 3-spaces (they tend to a Euclidean 3-space at spatial infinity). As a consequence each
3-space of the global non-inertial frame is a {\it non-inertial rest frame} of the 3-universe. At spatial infinity there are asymptotic inertial observers carrying a flat tetrad whose spatial axes are identified by the fixed stars of star catalogues.

\medskip
See Ref.\cite{18} for a review of this approach to GR with a rich bibliography.

\subsection{Tetrads and Cotetrads}

Assume that the world-line $x^{\mu}(\tau)$ of an arbitrary time-like
observer \footnote{An observer, or better a mathematical observer, is an idealization of a
measuring apparatus containing an atomic clock and defining, by means of
gyroscopes, a set of spatial axes (and then a, maybe orthonormal, tetrad
with a convention for its transport) in each point of the world-line.}
carrying a standard atomic clock is given: $\tau$ is an arbitrary
monotonically increasing function of the proper time of this clock. Then one
gives an admissible 3+1 splitting of the asymptotically flat space-time, namely a nice
foliation with space-like instantaneous 3-spaces $\Sigma_{\tau}$. It is the
mathematical idealization of a protocol for clock synchronization: all the
clocks in the points of $\Sigma_{\tau}$ sign the same time of the atomic
clock of the observer. The observer and the
foliation define a global non-inertial reference frame after a choice of
4-coordinates. On each 3-space $\Sigma_{\tau}$  one chooses curvilinear
3-coordinates $\sigma^r$ having the observer as origin.

The quantities $\sigma^A = (\tau; \sigma^r)$ are the Lorentz-scalar and observer-dependent
\textit{radar 4-coordinates}, first introduced by Bondi \cite{11}. \medskip

If $x^{\mu} \mapsto \sigma^A(x)$ is the coordinate transformation from
world 4-coordinates $x^{\mu}$ having the observer as origin to radar 4-coordinates, its inverse $\sigma^A
\mapsto x^{\mu} = z^{\mu}(\tau ,\sigma^r)$ defines the \textit{embedding}
functions $z^{\mu}(\tau ,\sigma^r)$ describing the 3-spaces $\Sigma_{\tau}$
as embedded 3-manifolds into the asymptotically flat space-time.

Let $z^{\mu}_A\tau, \sigma^u) = \partial\, z^{\mu}(\tau, \sigma^u) / \partial\, \sigma^A$ denote the gradients of the
embedding functions with respect to the radar 4-coordinates.
The space-like 4-vectors $z^{\mu}_r(\tau ,\sigma^u)$ are tangent to $\Sigma_{\tau}$, so that the unit
time-like normal $l^{\mu}(\tau ,\sigma^u)$ is proportional to $\epsilon^{\mu}{}_{%
\alpha \beta\gamma}\, [z^{\alpha}_1\, z^{\beta}_2\, z^{\gamma}_3](\tau
,\sigma^u)$ ($\epsilon_{\mu\alpha\beta\gamma}$ is the Levi-Civita tensor). Instead $z^{\mu}_{\tau}(\tau, \sigma^u)$ is a time-like 4-vector skew with respect to the 3-spaces leaves of the foliation
\footnote{In special relativity, see Refs. \cite{8,9,10},
one has $z^{\mu}_{\tau}(\tau ,\sigma^r) = [N\, l^{\mu} + N^r\,
z^{\mu}_r](\tau ,\sigma^r)$ with $N(\tau ,\sigma^r) = \epsilon\,
[z^{\mu}_{\tau}\, l_{\mu}](\tau ,\sigma^r) = 1 + n(\tau, \sigma^r) > 0$ and $
N_r(\tau ,\sigma^r) = - \epsilon\, [z^{\mu}_{\tau}\, \eta_{\mu\nu}\, z_r^{\mu}](\tau ,\sigma^r)$ being the
lapse and shift functions respectively of the global non-inertial frame of Minkowski space-time so defined.}.\medskip

In GR the dynamical fields are the components ${}^4g_{\mu\nu}(x)$ of
the 4-metric and not the  embeddings $x^{\mu} = z^{\mu}(\tau,
\sigma^r)$ defining the admissible 3+1 splittings of space-time like
in  the parametrized Minkowski theories of special relativity \cite{9,10}. Now the gradients
$z^{\mu}_A(\tau, \sigma^r)$ of the embeddings give the transition
coefficients from radar to world 4-coordinates, so that the
components ${}^4g_{AB}(\tau, \sigma^r) = z^{\mu}_A(\tau, \sigma^r)\,
z^{\nu}_B(\tau, \sigma^r)\, {}^4g_{\mu\nu}(z(\tau, \sigma^r))$ of
the 4-metric will be the dynamical fields in the ADM action \cite{1}.
\medskip

Let us remark that {\it the ten quantities ${}^4g_{AB}(\tau, \sigma^r)$ are
4-scalars of the space-time due to the use of the Lorentz-scalar radar 4-coordinates}. In each
3-space $\Sigma_{\tau}$ considered as a 3-manifold with 3-coordinates $\sigma^r$ (and not as a 3-sub-manifold of the space-time) ${}^4g_{\tau r}(\tau, \sigma^u)$ is a 3-vector and ${}^4g_{rs}(\tau, \sigma^u)$ is a 3-tensor.

Therefore {\it all the components of  "radar tensors", i.e. tensors expressed in radar 4-coordinates, are
4-scalars of the space-time}.
\medskip

The 4-metric ${}^4g_{AB}$ has signature $\sgn\, (+---)$ with $\sgn =
\pm$ (the particle physics, $\sgn = +$, and general relativity,
$\sgn = -$, conventions). Flat indices $(\alpha )$, $\alpha = o, a$,
are raised and lowered by the flat Minkowski metric
${}^4\eta_{(\alpha )(\beta )} = \sgn\, (+---)$. We define
${}^4\eta_{(a)(b)} = - \sgn\, \delta_{(a)(b)}$ with a
positive-definite Euclidean 3-metric. From now on we shall denote
the curvilinear 3-coordinates $\sigma^r$ with the notation $\vec
\sigma$ for the sake of simplicity. Usually the convention of sum
over repeated indices is used, except when there are too many
summations. The symbol $\approx$ means Dirac weak equality, while the symbol
$\cir$ means evaluated by using the equations of motion. For the curvature tensors
we use the conventions of Misner-Thorne-Wheeler \cite{19}, which has $\sgn = - 1$, with a minus sign with
respect to Wald \cite{20}.

\bigskip

We shall work with the tetrads ${}^4E^A_{(\alpha)}(\tau, \vec
\sigma)$ and the cotetrads ${}^4E_A^{(\alpha)}(\tau, \vec \sigma)$
\cite{7,20} ($(\alpha )$ are flat indices). To rebuild the original tetrads
${}^4E^{\mu}_{(\alpha)}(\tau, \vec \sigma) = z^{\mu}_A(\tau, \vec
\sigma)\, {}^4E^A_{(\alpha)}(\tau, \vec \sigma)$ we must know
explicitly the embedding $z^{\mu}(\tau ,\vec \sigma)$ of the
instantaneous 3-spaces, so to be able to evaluate the transformation
coefficients $ z^{\mu}_A(\tau, \vec \sigma)$.

\medskip

Since the world-line of the time-like observer can be chosen as the
origin of a set of the spatial world coordinates, i.e.
$x^{\mu}(\tau) = (x^o(\tau); 0)$, it turns out that with this choice
the space-like surfaces of constant coordinate time $x^o(\tau) =
const.$ coincide with the dynamical instantaneous 3-spaces
$\Sigma_{\tau}$ with $\tau = const.$. By using asymptotic flat
tetrads $\epsilon^{\mu}_A = \delta^{\mu}_o\, \delta^{\tau}_A +
\delta^{\mu}_i\, \delta^i_A$ (with $\epsilon^A_{\mu}$ denoting the
inverse flat cotetrads) and by choosing a coordinate world time
$x^o(\tau) = x^o_o + \epsilon^o_{\tau}\, \tau = x^o_o + \tau$, one
gets the following preferred embedding corresponding to these given
world 4-coordinates $x^{\mu} = z^{\mu}(\tau, \vec \sigma) =
x^{\mu}(\tau) + \epsilon^{\mu}_r\, \sigma^r = \delta^{\mu}_o\, x^o_o
+ \epsilon^{\mu}_A\, \sigma^A$. This choice implies $z^{\mu}_A(\tau,
\vec \sigma) = \epsilon^{\mu}_A$ and ${}^4g_{\mu\nu}(x = z(\tau,
\vec \sigma)) = \epsilon^A_{\mu}\, \epsilon_{\nu}^B\, {}^4g_{AB}(\tau,
\vec \sigma)$.

\bigskip

To take into account the coupling of fermions to the gravitational
field metric gravity has to be replaced with tetrad gravity. This
can be achieved by decomposing the 4-metric on cotetrad fields

\beq
 {}^4g_{AB}(\tau, \vec \sigma) = E_A^{(\alpha)}(\tau, \vec \sigma)\,
 {}^4\eta_{(\alpha)(\beta)}\, E^{(\beta)}_B(\tau, \vec \sigma),
 \label{2.1}
 \eeq

\noindent by putting this expression into the ADM action and by
considering the resulting action, a functional of the 16 fields
$E^{(\alpha)}_A(\tau, \vec \sigma)$, as the action for ADM tetrad
gravity.

\medskip

This leads to an interpretation of gravity based on a congruence of
time-like observers endowed with orthonormal tetrads: in each point
of space-time the time-like axis is the  unit 4-velocity of the
observer, while the spatial axes are a (gauge) convention for
observer's gyroscopes. This framework was developed in  Refs.\cite{8,19}.
\medskip

General tetrads ${}^4E^A_{(\alpha )}(\tau, \vec \sigma)$ and
cotetrads ${}^4E_A^{(\alpha )}(\tau, \vec \sigma)$ are connected to
the tetrads $\eo^A_{(\beta )}$ and cotetrads $\eo_A^{(\beta )}$
adapted to the 3+1 splitting (the time-like tetrad is the unit
normal $l^A$ to $\Sigma_{\tau}$; this choice corresponds to the so
called {\it Schwinger time gauges}) by a point-dependent standard
Lorentz boost for time-like orbits with boost parameters $\varphi_{(a)}(\tau, \vec \sigma)$
 acting on the flat indices
\footnote{As a consequence, the flat indices
$(a)$ of the adapted tetrads and cotetrads and of the triads and
cotriads on $\Sigma_{\tau}$ transform as Wigner spin-1 indices under
the point-dependent SO(3) Wigner rotations associated with Lorentz
transformations in the tangent
plane to the space-time in the given point of $\Sigma_{\tau}$.
Instead the index $(o)$ of the adapted tetrads and cotetrads is a
local Lorentz scalar index.}

\medskip

\bea
  {}^4E^A_{(\alpha )} &=& \eo^A_{(\beta )}\, L^{(\beta )}{}_{(\alpha
 )}(\varphi_{(a)}),\qquad
 {}^4E^{(\alpha )}_A = L^{(\alpha )}{}_{(\beta )}(\varphi_{(a)})\, \eo^{(\beta
 )}_A, \nonumber \\
 &&{}\nonumber \\
 {}^4g_{AB} &=& {}^4E^{(\alpha )}_A\, {}^4\eta_{(\alpha
 )(\beta )}\, {}^4E^{(\beta )}_B = \eo^{(\alpha )}_A\, {}^4\eta_{(\alpha
 )(\beta )}\, \eo^{(\beta )}_B.
 \label{2.2}
 \eea

The adapted tetrads and cotetrads  have the expression

\bea
 \eo^A_{(o)} &=& {1\over {1 + n}}\, (1; - n_{(a)}\,
 {}^3e^r_{(a)}) = l^A,\qquad \eo^A_{(a)} = (0; {}^3e^r_{(a)}), \nonumber \\
 &&{}\nonumber  \\
 \eo^{(o)}_A &=& (1 + n)\, (1; \vec 0) = \sgn\, l_A,\qquad \eo^{(a)}_A
= (n_{(a)}; {}^3e_{(a)r}),
 \label{2.3}
 \eea

\noindent where ${}^3e^r_{(a)}$ and ${}^3e_{(a)r}$ are triads and
cotriads on $\Sigma_{\tau}$, $N = 1 + n > 0$ is the lapse function and $n_{(a)} = n_r\, {}^3e^r_{(a)} =
n^r\, {}^3e_{(a)r}$ \footnote{Since we use the positive-definite
3-metric $\delta_{(a)(b)} $, we shall use only lower flat spatial
indices. Therefore for the cotriads we use the notation
${}^3e^{(a)}_r\,\, {\buildrel {def}\over =}\, {}^3e_{(a)r}$ with
$\delta_{(a)(b)} = {}^3e^r_{(a)}\, {}^3e_{(b)r}$.} are adapted shift
functions. Both $n$ and $n_{(a)}$ are 4-scalars of the space-time.

\bigskip

The adapted tetrads $\eo^A_{(a)}$ are defined modulo SO(3) rotations
$\eo^A_{(a)} = R_{(a)(b)}(\alpha_{(e)})\, {}^4{\buildrel \circ \over
{\bar E}}^A_{(b)}$, ${}^3e^r_{(a)} = R_{(a)(b)}(\alpha_{(e)})\,
{}^3{\bar e}^r_{(b)}$, where $\alpha_{(a)}(\tau ,\vec \sigma )$ are
three point-dependent Euler angles. After having chosen an arbitrary
point-dependent origin $\alpha_{(a)}(\tau ,\vec \sigma ) = 0$, we
arrive at the following adapted tetrads and cotetrads [${\bar
n}_{(a)} = \sum_b\, n_{(b)}\, R_{(b)(a)}(\alpha_{(e)})\,$]

\bea
 {}^4{\buildrel \circ \over {\bar E}}^A_{(o)}
 &=& \eo^A_{(o)} = {1\over {1 + n}}\, (1; - {\bar n}_{(a)}\,
 {}^3{\bar e}^r_{(a)}) = l^A,\qquad {}^4{\buildrel \circ \over
 {\bar E}}^A_{(a)} = (0; {}^3{\bar e}^r_{(a)}), \nonumber \\
 &&{}\nonumber  \\
 {}^4{\buildrel \circ \over {\bar E}}^{(o)}_A
 &=& \eo^{(o)}_A = (1 + n)\, (1; \vec 0) = \sgn\, l_A,\qquad
 {}^4{\buildrel \circ \over {\bar E}}^{(a)}_A
= ({\bar n}_{(a)}; {}^3{\bar e}_{(a)r}),
 \label{2.4}
 \eea

\noindent which we shall use as a reference standard.\medskip

Then Eqs.(\ref{2.2}), namely

\beq
 {}^4E^A_{(\alpha )} = {}^4{\buildrel \circ \over {\bar E}}^A_{(o)}\,
 L^{(o)}{}_{(\alpha )}(\varphi_{(c)}) + {}^4{\buildrel \circ \over
 {\bar E}}^A_{(b)}\, R^T_{(b)(a)}(\alpha_{(c)})\,
 L^{(a)}{}_{(\alpha )}(\varphi_{(c)}),
 \label{2.5}
 \eeq
\medskip

\noindent show that every point-dependent Lorentz transformation
 $\Lambda$ in the tangent planes may be parametrized with the
 (Wigner) boost parameters $\varphi_{(a)}$ and the Euler angles
 $\alpha_{(a)}$, being the product $\Lambda = R\, L$ of a rotation
 and a boost.

\bigskip

The future-oriented unit normal to $\Sigma_{\tau}$ is\medskip

\bea
 l_A &=& \sgn\, (1 + n)\, \Big(1;\, 0\Big),\qquad {}^4g^{AB}\, l_A\, l_B =
\sgn ,\nonumber \\
 &&{}\nonumber \\
 l^A &=& \sgn\, (1 + n)\, {}^4g^{A\tau} = {1\over {1 + n}}\, \Big(1;\, - n^r\Big) =
{1\over {1 + n}}\, \Big(1;\, - {\bar n}_{(a)}\, {}^3{\bar e}_{(a)}^r\Big).\nonumber \\
 &&{}
 \label{2.6}
 \eea

\subsection{The 4-metric and the Canonical Variables.}

The 4-metric has the following expression

 \bea
 {}^4g_{\tau\tau} &=& \sgn\, [(1 + n)^2 - {}^3g^{rs}\, n_r\,
 n_s] = \sgn\, [(1 + n)^2 - {\bar n}_{(a)}\, {\bar n}_{(a)}],\nonumber \\
 {}^4g_{\tau r} &=& - \sgn\, n_r = -\sgn\, {\bar n}_{(a)}\,
 {}^3{\bar e}_{(a)r},\nonumber \\
  {}^4g_{rs} &=& -\sgn\, {}^3g_{rs},\qquad
 {}^3g_{rs} = {}^3{\bar e}_{(a)r}\, {}^3{\bar e}_{(a)s},\qquad {}^3g^{rs} =
 {}^3{\bar e}^r_{(a)}\, {}^3{\bar e}^s_{(a)},\nonumber \\
 &&{}\nonumber \\
 {}^4g^{\tau\tau} &=& {{\sgn}\over {(1 + n)^2}},\qquad
  {}^4g^{\tau r} = -\sgn\, {{n^r}\over {(1 + n)^2}} = -\sgn\, {{{}^3{\bar e}^r_{(a)}\,
 {\bar n}_{(a)}}\over {(1 + n)^2}},\nonumber \\
 {}^4g^{rs} &=& -\sgn\, ({}^3g^{rs} - {{n^r\, n^s}\over
 {(1 + n)^2}}) = -\sgn\, {}^3{\bar e}^r_{(a)}\, {}^3{\bar e}^s_{(b)}\, (\delta_{(a)(b)} -
 {{{\bar n}_{(a)}\, {\bar n}_{(b)}}\over {(1 + n)^2}}),\nonumber \\
 &&{}\nonumber \\
 &&{}^3g = \gamma =
({}^3e)^2,\quad {}^3e = det\, {}^3{\bar e}_{(a)r},\quad
 \sqrt{|{}^4g|} = {{\sqrt{{}^3g}}\over {\sqrt{\sgn\,
{}^4g^{\tau\tau}}}} = \sqrt{\gamma}\, (1 + n) = {}^3e\, (1 + n).\nonumber \\
&&{}
 \label{2.7}
 \eea

\bigskip

The 3-metric ${}^3g_{rs}$ has signature $(+++)$, so that we
may put all the flat 3-indices {\it down}. We have ${}^3g^{ru}\,
{}^3g_{us} = \delta^r_s$,
$\partial_A\, {}^3g^{rs} = - {}^3g^{ru}\, {}^3g^{sv}\, \partial_A\,
{}^3g_{uv}$. The shift functions $n_r$ are 3-vectors of the 3-spaces $\Sigma_{\tau}$
and 4-scalars (like ${\bar n}_{(a)}$) of the space-time.\bigskip

The conditions for having an admissible 3+1 splitting of space-time
are \cite{9,14}:\medskip

a) $1 + n(\tau ,\vec \sigma) > 0$ everywhere (the instantaneous
3-spaces never intersect each other);\hfill\break

b) $\sgn\, {}^4g_{\tau\tau}(\tau ,\vec \sigma) > 0$, i.e. $(1 + n(\tau ,\vec \sigma))^2 > [{}^3g^{rs}\,
n_r\, n_s](\tau ,\vec \sigma)$ (the rotational velocity never exceeds the velocity of
light $c$, so that the coordinate singularity of the rotating disk
named "horizon problem" is avoided); \hfill\break

c) $\sgn\, {}^4g_{rr}(\tau ,\vec \sigma) = - {}^3g_{rr}(\tau ,\vec \sigma) < 0$ (satisfied by the
signature of ${}^3g_{rs}$), $[{}^4g_{rr}\, {}^4g_{ss} -
({}^4g_{rs})^2](\tau ,\vec \sigma) > 0$ and $ det\, \sgn\, {}^4g_{rs}(\tau ,\vec \sigma) = - det\,
{}^3g_{rs}(\tau ,\vec \sigma) < 0$ (satisfied by the signature of ${}^3g_{rs}$) so that
$det\, {}^4g_{AB}(\tau ,\vec \sigma) < 0$; these conditions imply that ${}^3g_{rs}(\tau ,\vec \sigma)$
 has three definite positive eigenvalues $\lambda_r(\tau ,\vec \sigma) =
\Lambda_r^2(\tau ,\vec \sigma)$;\hfill\break

d) the space-time is asymptotically Minkowskian with the
instantaneous 3-spaces orthogonal to the ADM 4-momentum at spatial
infinity: they are non-inertial rest frames of the 3-universe
(isolated system), there is an asymptotic Minkowski background
4-metric and there are asymptotic inertial observers whose spatial
axes $\epsilon^{\mu}_r$ are  identified by the fixed stars.

\bigskip

As said in Ref.\cite{12}, in ADM canonical tetrad gravity the 16
configuration variables are: the 3 boost variables
$\varphi_{(a)}(\tau, \vec \sigma)$; the lapse and shift functions
$n(\tau, \vec \sigma)$ and $n_{(a)}(\tau, \vec \sigma)$; the
cotriads ${}^3e_{(a)r}(\tau, \vec \sigma)$. Their conjugate momenta
are $\pi_{\varphi_{(a)}}(\tau, \vec \sigma)$, $\pi_n(\tau, \vec
\sigma)$, $\pi_{n_{(a)}}(\tau, \vec \sigma)$, ${}^3\pi^r_{(a)}(\tau,
\vec \sigma)$. There are 14 first-class constraints: A) the 10
primary constraints $\pi_{\varphi_{(a)}}(\tau, \vec \sigma) \approx
0$, $\pi_n(\tau, \vec \sigma) \approx 0$, $\pi_{n_{(a)}}(\tau, \vec
\sigma) \approx 0$ and the 3 rotation constraints $M_{(a)}(\tau,
\vec \sigma) \approx 0$ implying the gauge nature of the 3 Euler
angles $\alpha_{(a)}(\tau, \vec \sigma)$; B) the 4 secondary
super-Hamiltonian and super-momentum constraints ${\cal H}(\tau,
\vec \sigma) \approx 0$, ${\cal H}_{(a)}(\tau, \vec \sigma) \approx
0$. As a consequence there are 14 gauge variables (the inertial
effects) and two pairs of canonically conjugate physical degrees of
freedom (the tidal effects). At this stage the basis of canonical variables for this formulation
of tetrad gravity, naturally adapted to 7 of the 14 first-class
constraints, is

\beq
 \begin{minipage}[t]{3cm}
\begin{tabular}{|l|l|l|l|} \hline
$\varphi_{(a)}$ & $n$ & $n_{(a)}$ & ${}^3e_{(a)r}$ \\ \hline $
\pi_{\varphi_{(a)}}\, \approx 0$ & $\pi_n\, \approx 0$ &
$\pi_{n_{(a)}}\, \approx 0 $ & ${}^3{ \pi}^r_{(a)}$
\\ \hline
\end{tabular}
\end{minipage}
 \label{2.8}
 \eeq

From Eqs.(5.5) of Ref.\cite{21} we assume the following
(direction-independent, so to kill super-translations) boundary
conditions at spatial infinity ($r = \sqrt{\sum_r\, (\sigma^r)^2}$;
$\epsilon > 0$; $M = const.$): $n(\tau, \vec \sigma)
\rightarrow_{r\, \rightarrow\, \infty}\, O(r^{- (2 + \epsilon)})$,
$\pi_n(\tau, \vec \sigma) \rightarrow_{r\, \rightarrow\, \infty}\,
O(r^{-3})$, $n_{(a)}(\tau, \vec \sigma) \rightarrow_{r\,
\rightarrow\, \infty}\, O(r^{- \epsilon})$, $\pi_{n_{(a)}}(\tau,
\vec \sigma) \rightarrow_{r\, \rightarrow\, \infty}\, O(r^{-3})$,
$\varphi_{(a)}(\tau, \vec \sigma) \rightarrow_{r\, \rightarrow\,
\infty}\, O(r^{- (1 + \epsilon)})$, $\pi_{\varphi_{(a)}}(\tau, \vec
\sigma) \rightarrow_{r\, \rightarrow\, \infty}\, O(r^{-2})$,
${}^3e_{(a)r}(\tau, \vec \sigma) \rightarrow_{r\, \rightarrow\,
\infty}\, \Big(1 + {M\over {2 r}}\Big)\, \delta_{ar} + O(r^{-
3/2})$, ${}^3\pi^r_{(a)}(\tau, \vec \sigma) \rightarrow_{r\,
\rightarrow\, \infty}\, O(r^{- 5/2})$.

\subsection{The York Canonical Basis.}

In Ref.\cite{12} we studied the following point canonical
transformation (it is a Shanmugadhasan canonical transformation
\cite{3}) on the canonical variables (\ref{2.8}), implementing the
York map of Refs.\cite{13,22} and identifying a canonical basis
adapted to the 10 primary first-class constraints . It leads to the following York canonical basis

\bea
 &&\begin{minipage}[t]{3cm}
\begin{tabular}{|l|l|l|l|} \hline
$\varphi_{(a)}$ & $n$ & $n_{(a)}$ & ${}^3e_{(a)r}$ \\ \hline
$\pi_{\varphi_{(a)}} \approx 0$ & $\pi_n \approx 0$ & $
\pi_{n_{(a)}} \approx 0 $ & ${}^3{ \pi}^r_{(a)}$
\\ \hline
\end{tabular}
\end{minipage} \hspace{1cm}\nonumber \\
 &&{}\nonumber \\
 &&{\longrightarrow \hspace{.2cm}} \
\begin{minipage}[t]{4 cm}
\begin{tabular}{|ll|ll|l|l|l|} \hline
$\varphi_{(a)}$ & $\alpha_{(a)}$ & $n$ & ${\bar n}_{(a)}$ &
$\theta^r$ & $\tilde \phi$ & $R_{\bar a}$\\ \hline
$\pi_{\varphi_{(a)}} \approx0$ &
 $\pi^{(\alpha)}_{(a)} \approx 0$ & $\pi_n \approx 0$ & $\pi_{{\bar n}_{(a)}} \approx 0$
& $\pi^{(\theta )}_r$ & $\pi_{\tilde \phi}$ & $\Pi_{\bar a}$ \\
\hline
\end{tabular}
\end{minipage}\nonumber \\
 &&{}
 \label{2.9}
 \eea

\noindent where ${}^3{\bar e}_{(a)r} = \sum_b\, {}^3e_{(b)r}\,
R_{(b)(a)}(\alpha_{(e)})$ (with conjugate momenta  ${}^3{\bar
\pi}^r_{(a)}$), ${\bar n}_{(a)} = \sum_b\, n_{(b)}\,
R_{(b)(a)}(\alpha_{(e)})$ are the cotriads and the shift functions
at $\alpha_{(a)}(\tau ,\vec \sigma ) = 0$ after the extraction of
the rotation matrix $R_{(a)(b)}(\alpha_{(e)}(\tau ,\vec \sigma))$,
see after Eq.(\ref{2.3}). Due to the use of radar 4-coordinates all
the canonical variables of the York basis are 4-scalars of the space-time,
but they have different 3-tensorial behaviors inside the 3-spaces. $\theta^i$ and $\pi_{\tilde \phi}$
are the primary inertial gauge variables, while $n$ and ${\bar n}_{(a)}$ are the secondary ones.
\bigskip

Since the 3-metric $ {}^3g_{rs}$ is a real symmetric $3 \times 3$
matrix, it may be diagonalized with an {\it orthogonal} matrix
$V(\theta^r)$, $V^{-1} = V^T$ ($\sum_u\, V_{ua}\, V_{ub} =
\delta_{ab}$, $\sum_a\, V_{ua}\, V_{va} = \delta_{uv}$, $\sum_{uv}\,
\epsilon_{wuv}\, V_{ua}\, V_{vb} = \sum_c\, \epsilon_{abc}\,
V_{cw}$), $det\, V = 1$, depending on 3 Euler angles $\theta^r$
\footnote{Due to the positive signature of the 3-metric, we define
the matrix $V$ with the following indices: $V_{ru}$. Since the
choice of Shanmugadhasan canonical bases breaks manifest covariance,
we will use the notation $V_{ua} = \sum_v\, V_{uv}\, \delta_{v(a)}$
instead of $V_{u(a)}$. We use the following types of indices: $a =
1,2,3$ and $\bar a = 1,2$.}. The gauge Euler angles $\theta^r$ give
a  description of the 3-coordinate systems on $\Sigma_{\tau}$ from a
local point of view, because they give the orientation of the
tangents to the three 3-coordinate lines through each point (their
conjugate momenta are determined by the super-momentum constraints).
However it is more convenient to choose the three gauge parameters as first
kind coordinates $\theta^i(\tau, \vec \sigma)$ ($- \infty < \theta^i
< + \infty$) on the O(3) group manifold, so that by definition we
have $A_{ri}(\theta^n)\, \theta^i = \delta_{ri}\, \theta^i$. In this
case we have $V_{ru}(\theta^i) = \Big(e^{- \sum_i\, {\hat T}_i\,
\theta^i}\Big)_{ru}$, where $({\hat T}_i)_{ru} = \epsilon_{rui}$ are
the generators of the o(3) Lie algebra in the adjoint
representation, and the Euler angles may be expressed as ${\hat
\theta}^i = f^i(\theta^n)$. Since the Cartan matrix has the form
$A(\theta^n) = {{1 - e^{- \sum_i\, {\hat T}_i\, \theta^i} }\over
{\sum_i\, {\hat T}_i\, \theta^i}}$, we get the following expansions
around $\theta^i = 0$: $V_{ru}(\theta^i)\, \rightarrow_{\theta^i
\rightarrow 0}\, \delta_{ru} - \epsilon_{rui}\, \theta^i +
O(\theta^2)$, $A_{ru}(\theta^i)\, \rightarrow_{\theta^i \rightarrow
0}\, \delta_{ru} - {1 \over 2}\, \epsilon_{rui}\, \theta^i +
O(\theta^2)$, $B_{ru}(\theta^i)\, \rightarrow_{\theta^i \rightarrow
0}\, \delta_{ru} + {1\over 2}\, \epsilon_{rui}\, \theta^i +
O(\theta^2)$.

\bigskip

In the York canonical basis we have (from now on we will use
$V_{ra}$ for $V_{ra}(\theta^n)$ to simplify the notation)

\begin{eqnarray*}
 {}^3e_{(a)r} &=& \sum_b\, R_{(a)(b)}(\alpha_{(e)})\, {}^3{\bar
 e}_{(b)r},\qquad {}^3{\bar e}_{(a)r} = {\tilde \phi}^{1/3}\, Q_a\,
 V_{ra},\nonumber \\
 {}^3e^r_{(a)} &=& \sum_b\, R_{(a)(b)}(\alpha_{(e)})\, {}^3{\bar
 e}^r_{(b)},\qquad {}^3{\bar e}^r_{(a)} = {\tilde \phi}^{- 1/3}\, Q^{-1}_a\,
 V_{ra},
 \end{eqnarray*}

\begin{eqnarray*}
 {}^4g_{\tau\tau} &=& \sgn\, \Big[(1 + n)^2 - \sum_a\,
 {\bar n}_{(a)}^2\Big],\nonumber \\
 {}^4g_{\tau r} &=& - \sgn\, \sum_a\, {\bar n}_{(a)}\, {}^3{\bar e}_{(a)r} =
 - \sgn\, {\tilde \phi}^{1/3}\, \sum_a\, Q_a\,
 V_{ra}\, {\bar n}_{(a)},\nonumber \\
 {}^4g_{rs} &=& - \sgn\, {}^3g_{rs}
 = - \sgn\, \sum_{uv}\, V_{ru}\, \lambda_u\, \delta_{uv}\,
 V^T_{vs} = - \sgn\, \sum_a\, \Big(V_{ra}\, \Lambda^a\Big)\,
 \Big(V_{sa}\, \Lambda^a\Big) =\nonumber \\
 &=& - \sgn\, \sum_a\, {}^3{\bar e}_{(a)r}\, {}^3{\bar e}_{(a)s} =
 - \sgn\, \phi^4\, {}^3{\hat g}_{rs}\, =\, - \sgn\, {\tilde \phi}^{2/3}\,
 \sum_a\, Q^2_a\, V_{ra}\, V_{sa},\nonumber \\
 &&{}\nonumber \\
 \Lambda_a &=& \sum_u\, \delta_{au}\, \sqrt{\lambda_u}\,\,
 =\,\, \phi^2\, Q_a\,  =\, {\tilde \phi}^{1/3}\, Q_a,\qquad
 Q_a\, =\, e^{\sum_{\bar a}^{1,2}\, \gamma_{\bar aa}\, R_{\bar a}}
 = e^{\Gamma_a^{(1)}}, \nonumber \\
 &&\sum_a\, \Gamma_a^{(1)} = 0,\qquad R_{\bar a} = \sum_a\,
 \gamma_{\bar aa}\, \Gamma_a^{(1)},\nonumber \\
 \tilde \phi &=& \phi^6 = \sqrt{\gamma} =
 \sqrt{det\, {}^3g} = {}^3\bar e = \Lambda_1\, \Lambda_2\, \Lambda_3,
 \qquad {}^3g^{rs} = {\tilde \phi}^{-2/3}\,
 \sum_a\, Q^{-2}_a\, V_{ra}\, V_{sa}, \nonumber \\
  \end{eqnarray*}

\begin{eqnarray*}
 {}^4g^{\tau\tau} &=& {{\sgn}\over {(1 + n)^2}},\qquad
  {}^4g^{\tau r} = -\sgn\, {\tilde \phi}^{-1/3}\,  {{Q_a^{-1}\, V_{ra}\,
  {\bar n}_{(a)}}\over {(1 + n)^2}},\nonumber \\
 {}^4g^{rs} &=&  -\sgn\, {\tilde \phi}^{-2/3}\, Q_a^{-1}\,
 Q_b^{-1}\, V_{ra}\, V_{sb}\, (\delta_{(a)(b)} -
 {{{\bar n}_{(a)}\, {\bar n}_{(b)}}\over {(1 + n)^2}}),
 \end{eqnarray*}

 \begin{eqnarray*}
 {}^3\pi^r_{(a)} &=& \sum_b\,
 R_{(a)(b)}(\alpha_{(e)})\, {\bar \pi}^r_{(b)} =
  \sum_b\, R_{(a)(b)}(\alpha_{(e)})\,
  {\tilde \phi}^{-1/3}\, \Big[
 V_{rb}\, Q^{-1}_b\, (\tilde \phi\, \pi_{\tilde \phi} +  \sum_{\bar b}\,
 \gamma_{\bar bb}\, \Pi_{\bar b}) +\nonumber \\
 &+& \sum_{l}^{l \not= b}\, \sum_{twi}\, Q^{-1}_l\, {{V_{rl}\,
 \epsilon_{blt}\, V_{wt}}\over {Q_l\, Q^{-1}_b - Q_b\, Q^{-1}_l
}}\, B_{iw}\, \pi^{(\theta )}_i \Big],
 \end{eqnarray*}

\bea
 \pi^{(\theta)}_i &=& - \sum_{lmrab}\, A_{ml}(\theta^n)\,
 \epsilon_{mir}\,  R_{(a)(b)}(\alpha_{(e)})\,
  {}^3{\bar e}_{(b)l}\, {}^3{\bar \pi}^r_{(a)},\nonumber \\
  \pi_{\tilde \phi} &=&   {{c^3}\over {12\pi\, G}}\, {}^3K
  \approx {1\over {3\,\, {}^3e}}\, \sum_{rab}\, {}^3{\bar
  \pi}^r_{(a)}\, R_{(a)(b)}(\alpha_{(e)})\, {}^3{\bar e}_{(b)r},
  \nonumber \\
  \Pi_{\bar a} &=& \sum_{rab}\, \gamma_{\bar ab}\, {}^3{\bar
  \pi}^r_{(a)}\, R_{(a)(b)}(\alpha_{(e)})\, {}^3{\bar e}_{(b)r}.
 \label{2.10}
 \eea

\noindent The set of numerical parameters $\gamma_{\bar aa}$
satisfies \cite{23} $\sum_u\, \gamma_{\bar au} = 0$, $\sum_u\,
\gamma_{\bar a u}\, \gamma_{\bar b u} = \delta_{\bar a\bar b}$,
$\sum_{\bar a}\, \gamma_{\bar au}\, \gamma_{\bar av} = \delta_{uv} -
{1\over 3}$. Each solution of these equations defines a different
York canonical basis. The new boundary conditions are
$\Lambda_a(\tau ,\vec \sigma )\, = \Big({\tilde \phi}^{1/3}\,
Q_a\Big)(\tau, \vec \sigma)\, \rightarrow_{r \rightarrow \infty}\,\,
1 + {M\over {4r}} + {{a_a}\over {r^{3/2}}} + O(r^{-3})$, $\tilde
\phi (\tau ,\vec \sigma )\, \rightarrow_{r \rightarrow \infty}\,\, 1
+ O(r^{-1})$, $\pi_i^{(\theta )}(\tau ,\vec
\sigma )\, \rightarrow_{r \rightarrow \infty}\,\, O(r^{-4})$,
$\pi^{(\alpha )}_{(a)}(\tau ,\vec
\sigma )\, \rightarrow_{r \rightarrow \infty}\, O(r^{-5/2})$,
$\pi_{\tilde \phi}(\tau ,\vec \sigma
)\, \rightarrow_{r \rightarrow \infty}\,\, O(r^{-5/2})$.
The angles $\alpha_{(a)}(\tau ,\vec \sigma )$ and $\theta^i(\tau
,\vec \sigma )$ must tend to zero in a direction-independent way at
spatial infinity.

\medskip

In Eq.(\ref{2.10}) the quantity ${}^3K(\tau, \vec \sigma)$ is the
trace of the extrinsic curvature ${}^3K_{rs}(\tau, \vec \sigma)$ of
the instantaneous 3-spaces $\Sigma_{\tau}$. In the York canonical
basis the extrinsic curvature ${}^3K_{rs}$ has the following
expression \cite{12}

\bea
  {}^3K_{rs} &\approx&
  - {{4\pi\, G}\over {c^3}}\, {\tilde \phi}^{-1/3}\,
 \Big(\sum_a\, Q^2_a\, V_{ra}\, V_{sa}\, [2\, \sum_{\bar b}\, \gamma_{\bar ba}\,
 \Pi_{\bar b} -  \tilde \phi\, \pi_{\tilde \phi}] +\nonumber \\
 &+& \sum_{ab}\, Q_a\, Q_b\, (V_{ra}\, V_{sb} +
 V_{rb}\, V_{sa})\, \sum_{twi}\, {{\epsilon_{abt}\,
 V_{wt}\, B_{iw}\, \pi_i^{(\theta )}}\over {
 Q_b\, Q^{-1}_a  - Q_a\, Q^{-1}_b}} \Big) =\nonumber \\
 &=&  {\tilde \phi}^{2/3}\, \Big[ {{4\pi\, G}\over
 {c^3}}\, \sum_a\, Q_a^2\, V_{ra}\, V_{sa}\, \Big( \pi_{\tilde \phi}\,
 -2\, {\tilde \phi}^{-1}\, \sum_{\bar a}\, \gamma_{\bar aa}\, \Pi_{\bar a}\Big)\,
 +\nonumber \\
 &+& \sum_{ab}^{a \not= b}\, \sigma_{(a)(b)}\, Q_a\, Q_b\, V_{ra}\,
 V_{sb}\Big].
   \label{2.11}
 \eea

In the last line we introduced the expansion $\theta = - \epsilon\, {}^3K$ and the spatial components of the
cotetrad-adapted shear $\sigma_{(a)(b)} = \sigma_{AB}\, {}^4{\buildrel \circ \over {\bar E}}^A_{(a)}\,
{}^4{\buildrel \circ \over {\bar E}}^B_{(b )}$  of the Eulerian observers
\footnote{ The surface-forming congruence of the Eulerian observers is defined by using the unit normal to the 3-spaces of Eq.(\ref{2.6}) as 4-velocity field. One has $ \theta = {}^4\nabla_A\,\, l^A \theta = {}^4\nabla_A\,\, l^A$
and $\sigma_{AB} = \sigma_{BA} = - {{\sgn}\over 2}\, ({}^3a_A\,
 l_B + {}^3a_B\, l_A) + {{\sgn}\over 2}\, ({}^4\nabla_A\, l_B + {}^4\nabla_B\, l_A) -
 {1\over 3}\, \theta\, {}^3h_{AB}$, where ${}^3h_{AB} = {}^4g_{AB} - \epsilon\, l_A\, l_B$ is the projector onto the 3-space and ${}^3a = l^B\, {}^4\nabla_A\, l_B$ is the acceleration of the Eulerian observers.}
(see Refs.\cite{14,24} for more details). The components of the shear are $\sigma_{(a)(b)}{|}_{a \not= b} = - {{8\pi\, G}\over {c^3}}\, {\tilde \phi}^{-1}\, \sum_{tw}\, {{\epsilon_{abt}\, V_{wt}}\over
{Q_b\, Q_a^{-1} - Q_a\, Q_b^{-1}}}\, \sum_i\, B_{iw}\, \pi_i^{(\theta )}$ and
$\sigma_{(a)(a)} = - {{8\pi\, G}\over {c^3}}\, {\tilde \phi}^{-1}
 \sum_{\bar a}\, \gamma_{\bar aa}\, \Pi_{\bar a}$ ($\sum_a\, \sigma_{(a)(a)} = 0$).
\bigskip

$\alpha_{(a)}(\tau ,\vec \sigma )$ and $\varphi_{(a)}(\tau ,\vec
\sigma )$ are the 6 configuration variables parametrizing the O(3,1)
gauge freedom in the choice of the tetrads in the tangent plane to
each point of $\Sigma_{\tau}$ and describe the arbitrariness in the
choice of a tetrad to be associated to a time-like observer, whose
world-line goes through the point $(\tau ,\vec \sigma )$. They fix
{\it the unit 4-velocity of the observer and the conventions for the
orientation of gyroscopes and their transport along the world-line
of the observer}.\medskip

The York time ${}^3K(\tau ,\vec \sigma)$ is the only gauge variable among the momenta: this
is a reflex of the Lorentz signature of space-time, because
$\pi_{\tilde \phi}(\tau ,\vec \sigma)$ and $\theta^n(\tau ,\vec \sigma)$ can be used as a set of
4-coordinates \cite{7}. Its conjugate variable, to be determined by the super-Hamiltonian
constraint, is $\tilde \phi(\tau ,\vec \sigma) = \phi^6(\tau ,\vec \sigma) = {}^3\bar e(\tau ,\vec \sigma)$, which
is the {\it 3-volume density} on $\Sigma_{\tau}$: $V_R
= \int_R d^3\sigma\, \phi^6(\tau ,\vec \sigma)$, $R \subset \Sigma_{\tau}$. Since we
have ${}^3g_{rs}(\tau ,\vec \sigma) = [{\tilde \phi}^{2/3}\, {}^3{\hat g}_{rs}](\tau ,\vec \sigma)$ with
$det\, {}^3{\hat g}_{rs}(\tau ,\vec \sigma) = 1$, $\tilde \phi(\tau ,\vec \sigma)$ is also called the
conformal factor of the 3-metric.

\medskip

The two pairs of canonical variables $R_{\bar a}(\tau ,\vec \sigma)$, $\Pi_{\bar a}(\tau ,\vec \sigma) =
- {{c^3}\over {8 \pi G}}\, \tilde \phi(\tau ,\vec \sigma)\, \sum_a\, \gamma_{\bar aa}\, \sigma_{(a)(a)}(\tau ,\vec \sigma)$, $\bar a = 1,2$, describe the generalized {\it tidal effects}, namely
the independent degrees of freedom of the gravitational field. In
particular the configuration tidal variables $R_{\bar a}$ depend
{\it only on the eigenvalues of the 3-metric}. They are DO
 {\it only} with respect to the gauge transformations
generated by 10 of the 14 first class constraints. Let us remark
that, if we fix completely the gauge and we go to Dirac brackets,
then the only surviving dynamical variables $R_{\bar a}$ and
$\Pi_{\bar a}$ become two pairs of {\it non canonical} DO for that gauge.
 \medskip

The gauge variables $\theta^i(\tau, \vec \sigma)$, $n(\tau, \vec
\sigma)$, ${\bar n}_{(a)}(\tau, \vec \sigma)$ describe inertial
effects, which are the the relativistic counterpart of the
non-relativistic ones (the centrifugal, Coriolis,... forces in
Newton mechanics in accelerated frames) and which are present also
in the non-inertial frames of Minkowski space-time \cite{9}. The unknowns in the super-momentum
constraints are the momenta $\pi_i^{(\theta)} = {{c^3}\over {8 \pi G}}\, \tilde \phi\,
\sum_{ab}\, \epsilon_{iab}\, Q_a\, Q_b^{-1}\, \sigma_{(a)(b)}{|}_{a \not= b}$; however, it is more convenient to solve these constraints in the shear components $\sigma_{(a)(b)}{|}_{a \not= b}$.

\bigskip

In the following we shall work in the {\it Schwinger time gauge}
$\varphi_{(a)}(\tau, \vec \sigma) \approx 0$ (tetrads adapted to the
3+1 splitting) and $\alpha_{(a)}(\tau,  \vec \sigma) \approx 0$
(arbitrary choice of an origin for rotations), where ${}^3e_{(a)r}(\tau ,\vec \sigma) \approx
{}^3{\bar e}_{(a)r}(\tau ,\vec \sigma)$.

\subsection{The 3-Geometry of the 3-Spaces}

In the 3-space $\Sigma_{\tau}$  the 3-Christoffel symbols have the
following expression \cite{14}

 \bea
   {}^3\Gamma^r_{uv}   &=& {1\over 2}\, {}^3g^{rs}\, \Big(\partial_u\, {}^3g_{vs}
 + \partial_v\, {}^3g_{us} - \partial_s\, {}^3g_{uv}\Big)
 =\nonumber \\
  &&{}\nonumber \\
   &=&{1\over 3}\,  (\delta_{ru}\, {\tilde \phi}^{-1}\, \partial_v\,
  \tilde \phi + \delta_{rv}\, {\tilde \phi}^{-1}\, \partial_u\, \tilde \phi ) -
   {1\over 3}\, \sum_{abs}\, Q^2_b\, Q^{-2}_a\, V_{ra}\, V_{sa}\,
  V_{ub}\, V_{vb}\, {\tilde \phi}^{-1}\, \partial_s\, \tilde \phi +\nonumber \\
  &+& \sum_{\bar aa}\, \gamma_{\bar aa}\, V_{ra}\, \Big(
  V_{ua}\, \partial_v\, R_{\bar a} + V_{va}\,
  \partial_u\, R_{\bar a}\Big) -
 \sum_{\bar aabs}\, \gamma_{\bar ab}\, Q^2_b\, Q^{-2}_a\,
  V_{ra}\, V_{sa}\,  V_{ub}\, V_{vb}\,
  \partial_s\, R_{\bar a} +\nonumber \\
  &+& {1\over 2}\, \sum_a\, V_{ra}\, \Big(\partial_u\, V_{va}
 + \partial_v\, V_{ua}\Big) +\nonumber \\
 &+& {1\over 2}\, \sum_{abs}\, Q_a^{-2}\, Q_b^2\, V_{ra}\,
 V_{sa}\, \Big[V_{ub}\, \Big(\partial_v\, V_{sb} -
 \partial_s\, V_{vb}\Big) +
  V_{vb}\, \Big(\partial_u\, V_{sb} - \partial_s\,
 V_{ub}\Big)\Big],\nonumber \\
 \sum_v\, {}^3\Gamma^v_{uv} &=&  {\tilde \phi}^{-1}\, \partial_u\,  \tilde
 \phi,
 \label{2.12}
 \eea

Eqs.(\ref{2.11}) and (\ref{2.12}) imply

 \bea
 {}^3K_{rs|u} &=& \partial_u\, {}^3K_{rs} -
 {}^3\Gamma^v_{ru}\, {}^3K_{vs} - {}^3\Gamma^v_{su}\, {}^3K_{rv}
 =\nonumber \\
 &=&{2\over 3}(\tilde \phi ^{-1} \partial_u \tilde \phi) {}^3K_{rs} - {}^3\Gamma_{ru}^v {}^3K_{vs} - {}^3\Gamma_{su}^v {}^3K_{rv} + \nonumber \\
&+& \Big[{4 \pi G \over c^3} \sum_a Q^2_a \Big( 2 (\partial_u \Gamma_a) V_{ra}V_{sa} (\pi_{\tilde \phi} - 2 \tilde \phi ^{-1} \sum_{\bar{a}} \gamma_{\bar{a}a}\Pi_{\bar{a}}) + \nonumber \\
&+& V_{ra} V_{sa} \Big[ \partial_u \pi_{\tilde \phi} - 2 \tilde \phi^{-1} \sum_{\bar{a}} \gamma_{\bar{a}a} ((\tilde \phi^{-1} \partial_u \tilde \phi) \Pi_{\bar{a}} -\partial_u \Pi_{\bar{a}}) \Big] +  \nonumber \\
&+& \partial_u (V_{ra}V_{sa} (\pi_{\tilde \phi} - 2 \tilde \phi^{-1} \sum_{\bar{a}} \gamma_{\bar{a}a} \Pi_{\bar{a}}))\Big) +  \nonumber \\
&+& \sum_{ab}^{a \not= b} Q_a Q_b V_{ra}V_{sb} \Big( \sigma_{(a)(b)} + \sigma_{(a)(b)} \partial_u(\Gamma_a + \Gamma_b) \Big) + \sum_{ab}^{a\not= b} \sigma_{(a)(b)} Q_a Q_b \partial_u (V_{ra}V_{sa})\Big] \nonumber \\
 {}&&
 \label{2.13}
 \eea

The quantity ${}^3K_{rs|u}$ will be needed for the Hamiltonian expression of the Riemann tensor.

\bigskip
See Appendix A for the lengthy Hamiltonian expression of
the 3-Riemann  and 3-Ricci tensors in the York basis.

\subsection{The First Half of Hamilton Equations}

In the York canonical basis the Dirac Hamiltonian  is

\bea
  H_D&=& {1\over c}\, {\hat E}_{ADM} + \int d^3\sigma\, \Big[ n\,
{\cal H} - {\bar n} _{(a)}\, {\bar {\cal H}}_{(a)}\Big](\tau ,
\sigma^u )
+ \lambda_r(\tau )\, {\hat P}^r_{ADM} +\nonumber \\
 &+&\int d^3\sigma\, \Big[\lambda_n\, \pi_n + \lambda_{
{\bar n}_{(a)}}\, \pi_{{\bar n}_{(a)}} + \lambda_{\varphi_{(a)}}\,
\pi_{ \varphi_{(a)}} + \lambda_{\alpha_(a)}\,
\pi^{(\alpha)}_{(a)}\Big](\tau , \sigma^u ),
 \label{2.14}
 \eea

\noindent with the weak ADM energy ${\hat E}_{ADM}$ given in Eqs. (3.45) and (B8) of
Ref.\cite{14}, where Eqs. (3.42) and (3.44) give the super-momentum
and super-Hamiltonian constraints respectively.\medskip

The $\lambda$'s are arbitrary Dirac multipliers: at the Hamiltonian
level they replace those Lagrangian velocities which are arbitrary
due to the gauge freedom. The Dirac multiplier $\lambda_r(\tau)$
implements the rest frame condition ${\hat P}^r_{ADM} \approx 0$.

With this Hamiltonian one gets the Hamilton equations (see Ref.\cite{14} for their explicit form),
whose content is equivalent to Einstein's equations. These equations are divided in five groups:

A) The {\it contracted Bianchi identities}, namely the evolution
equations for the solutions $\tilde \phi(\tau, \vec \sigma)$ and
$\pi_i^{(\theta)}(\tau, \vec \sigma)$ of the super-Hamiltonian and
super-momentum constraints: they are identities saying that, given a
solution of the constraints on a Cauchy surface, it remains a
solution also at later times.

B) The evolution equation for the four primary inertial gauge variables
$\theta^i(\tau , \vec \sigma)$ and ${}^3K(\tau , \vec \sigma)$ (the
equation for the York time is the Raychaudhuri equation): these equations
determine the lapse and the shift functions (the secondary inertial gauge variables) once four gauge-fixings
for the primary gauge variables are given.

C) The equations $\partial_{\tau}\, n(\tau , \vec \sigma) \cir
\lambda_n(\tau, \vec \sigma)$ and $\partial_{\tau}\, {\bar
n}_{(a)}(\tau , \vec \sigma) \cir \lambda_{{\bar n}_{(a)}}(\tau,
\vec \sigma)$. Once the lapse and shift functions of the chosen gauge
have been found, they determine the associated Dirac
multipliers.

D) The {\it hyperbolic} evolution partial differential equations (PDE) for the tidal variables
$R_{\bar a}(\tau , \vec \sigma)$, $\Pi_{\bar a}(\tau , \vec \sigma)$.

E) The Hamilton equations for matter, when present.

Given a solution of the super-momentum and super-Hamiltonian
constraints and the Cauchy data for the tidal variables on an
initial 3-space after having fixed the gauge, one can find a solution of Einstein's equations in
radar 4-coordinates adapted to a time-like observer in the chosen
gauge.

\medskip

To find the Hamiltonian expression of the 4-Christoffel symbols and of the 4-Riemann
tensor we only need the velocities $\partial_{\tau}\, {}^4g_{AB}$, namely only
the following three equations, belonging to the first half of Hamilton equations (the kinematical ones;  the equation for the 3-metric can also be obtained from Eq.(\ref{2.11}) for the extrinsic curvature) are

\bea
 \partial_{\tau}\, n(\tau , \vec \sigma ) &\cir& \lambda_n(\tau , \vec \sigma ),\nonumber \\
 \partial_{\tau}\, {\bar n}_{(a)}(\tau , \vec \sigma ) &\cir&
 \lambda_{\bar n(a)}(\tau , \vec \sigma),\nonumber \\
 {}&&\nonumber \\
   \partial_{\tau}\, {}^3g_{rs}(\tau, \vec \sigma)
 &=& \Big({\tilde \phi}^{2/3}\, \sum_a\, Q_a^2\, \Big[2\, ({1\over 3}\,
 {\tilde \phi}^{-1}\, \partial_{\tau}\, \tilde \phi + \partial_{\tau}\,
 \Gamma_a^{(1)})\, V_{ra}\, V_{sa} + \partial_{\tau}\, (V_{ra}\,
 V_{sa})\Big]\Big)(\tau, \vec \sigma).\nonumber \\
 &&{}
 \label{2.15}
 \end{eqnarray}

\subsection{Einstein Equations}

The 4-Ricci tensor is determined by Einstein equations (G is the Newton constant and
 the energy-momentum tensor is  defined as $T^{AB} =
- {2\over {\sqrt{|det {}^4g|}}}\, {{\delta S}\over {\delta\,
{}^4g_{AB}}}$)
\medskip

\bea
 {}^4G^{AB} &=& {}^4R^{AB} - {1\over 2}\, {}^4g^{AB}\, {}^4R\,\,\, \cir\,\,
 \,\,{{ 8\, \pi\, G}\over {c^3}}\, T^{AB},\nonumber \\
 &&{}\nonumber \\
 &&\Downarrow\nonumber \\
 &&{}\nonumber \\
 E_{AB} &{\buildrel {def}\over =}& {}^4R_{AB}\, - {{8\, \pi\,
 G}\over {c^3}}\, {\hat T}_{AB} \cir 0,\qquad {\hat T}_{AB}
 =T_{AB} - {1\over 2}\, {}^4g_{AB}\, T, \quad T = {}^4g^{AB}\, T_{AB}.
 \label{2.16}
 \eea

As shown in Refs.\cite{14,24} the Hamiltonian version of the matter
energy-momentum tensor $T^{AB}(\tau ,\vec \sigma )$ (depending upon
the 4-metric ${}^4g_{AB}(\tau ,\vec \sigma )$) in radar
4-coordinates on $\Sigma_{\tau}$ has:

1) $T^{\tau\tau} = {{{\cal M}}\over {{}^3e\, (1 + n)^2}}$ depending
upon the matter mass-energy density ${\cal M}(\tau ,\vec \sigma )$;

2) $T^{\tau r} = {1\over {{}^3e\, (1 + n)^2}}\, {}^3{\bar e}^r_{(a)}\,
\Big[(1 + n)\, {}^3{\bar e}^s_{(a)}\, {\cal M}_s - {\bar n}_{(a)}\, {\cal M}\Big]$
depending upon the matter mass-energy density and upon the matter
3-momentum density ${\cal M}_r(\tau ,\vec \sigma )$;

3) $T^{rs}$ with a form depending upon the given matter.

\bigskip

 The tetradic components of the energy-momentum tensor in the
 adapted basis (\ref{2.4}) are

 \bea
 T_{(\alpha )(\beta )} &=&
  {}^4{\buildrel \circ \over {\bar E}}^A_{(\alpha)}\,
  {}^4{\buildrel \circ \over {\bar E}}^B_{(\beta)}\, T_{AB},\nonumber \\
  &&T_{(o)(o)} = {1\over {(1 + n)^2}}\, \Big[T_{\tau\tau} - 2\,
  {\tilde \phi}^{-1/3}\, Q_a^{-1}\, {\bar n}_{(a)}\, T_{\tau r}\, V_{ra} +\nonumber \\
  &+&{\tilde \phi}^{-2/3}\, Q_a^{-1}\, Q_b^{-1}\, {\bar n}_{(a)}\, {\bar n}_{(b)}\,
  T_{rs}\, V_{ra}\, V_{sb}\Big],\nonumber \\
  &&T_{(o)(a)} = {{{\tilde \phi}^{-1/3}}\over {1 + n}}\,
  Q_a^{-1}\, V_{ra}\, \Big(T_{\tau r} - {\tilde \phi}^{-1/3}\,
  Q_b^{-1}\,  {\bar n}_{(b)}\,  T_{rs}\, V_{sb}\Big),\nonumber \\
  &&T_{(a)(b)} = {\tilde \phi}^{-2/3}\,  Q_a^{-1}\, Q_b^{-1}\, V_{ra}\,
 T_{rs}\, V_{sb}\,,\nonumber  \\
  &&{}\nonumber \\
  &&T_{rs} = {\tilde \phi}^{2/3}\,  Q_a^{1}\, Q_b^{1}\, V_{ra}\, V_{sb}\,
 T_{(a)(b)},\nonumber \\
  &&T_{\tau r} = {\tilde \phi}^{1/3}\, Q_a\, V_{ra}\, \Big((1 + n)\, T_{(o)(a)}
  + T_{(a)(b)}\,   {\bar n}_{(b)}\Big),\nonumber \\
  &&T_{\tau\tau} = (1 + n)^2\, T_{(o)(o)} + 2\, (1 + n)\,
  T_{(o)(a)}\,   {\bar n}_{(b)}  +\nonumber \\
  &+& T_{(a)(b)}\,  {\bar n}_{(a)}  {\bar n}_{(b)}.
 \label{2.17}
 \eea

\bigskip

Therefore the 4-Ricci tensor can be written in the following form:

\begin{eqnarray*}
 {}^4R_{\tau\tau} &=&
 E_{\tau\tau} + {{8\, \pi\, G}\over {c^3}}\,
 \Big[\Big(1 + {{{\bar n}_{(c)}\, {\bar n}_{(c)}}\over {(1 + n)^2}}\Big)\, {T_{\tau\tau}\over 2}
 -\nonumber \\
 &-& \Big(1 - {{{\bar n}_{(c)}\, {\bar n}_{(c)}}\over {(1 + n)^2}}\Big)\,
 \Big( {\tilde \phi}^{-1/3}\, {\bar n}_{(a)}\, Q_a^{-1}\, V_{ra}\, T_{\tau r}
 +\nonumber \\
 &+& {\tilde \phi}^{-2/3}\, ((1 + n)^2\,
 \delta_{(a)(b)})\, Q_a^{-1}\, Q_b^{-1}\, V_{ra}\, V_{sb}\, {T_{rs}\over 2}\Big)
 \Big],\nonumber \\
 {}^4R_{\tau r} &=&  E_{\tau r} + {{8\, \pi\, G}\over {c^3}}\, \Big[\Big(1 - {{{\bar n}_{(c)}\,
 {\bar n}_{(c)}}\over {(1 + n)^2}}\Big)\, T_{\tau r} -\nonumber \\
 &+& {\tilde \phi}^{1/3}\, {{{\bar n}_{(c)}\, Q_c\, V_{rc}}\over
 {2\, (1 + n)^2}}\, \Big(T_{\tau\tau} - {\tilde \phi}^{-2}\,
 ((1 + n)^2\, \delta_{(a)(b)} -\nonumber \\
 &-& {\bar n}_{(a)}\, {\bar n}_{(b)})\, Q_a^{-1}\, Q_b^{-1}\, V_{au}\,
 V_{vb}\, T_{uv}\Big) \Big],\nonumber \\
 {}^4R_{rs} &=&  E_{rs} + {{8\, \pi\, G}\over {c^3}}\,
  \Big[ {{{\tilde \phi}^{2/3}\, \sum_c\, Q_c^2\, V_{rc}\, V_{sc}}\over
  {2\, (1 + n)^2}}\, \Big(T_{\tau\tau} -\nonumber \\
 &-& 2\, {\tilde \phi}^{-1/3}\, {\bar n}_{(a)}\, Q_a^{-1}\,
 V_{ua}\, T_{\tau u}\Big) +  \Big(\delta^u_r\, \delta^v_s -
 {{ \sum_c\, Q_c^2\, V_{rc}\, V_{sc}}\over {2\, (1 + n)^2}}\, ((1 + n)^2\,
 \delta_{(a)(b)} -\nonumber \\
 &-& {\bar n}_{(a)}\, {\bar n}_{(b)})\, Q_a^{-1}\, Q_b^{-1}\, V_{ua}\,
 V_{vb}\Big)\, T_{uv}\Big],\nonumber \\
 \end{eqnarray*}

 \bea
  {}^4R\, &=&  {}^4g^{AB}\, E_{AB} - {{8\, \pi\, G}\over {c^3}}\,
 {{\sgn}\over {(1 + n)^2}}\, \Big[T_{\tau\tau} - 2\, {\tilde
 \phi}^{-1/3}\, {\bar n}_{(a)}\, Q_a^{-1}\, V_{ra}\, T_{\tau r} -\nonumber \\
 &-& 9\, {\tilde \phi}^{-2/3}\, ((1 + n)^2\, \delta_{(a)(b)}
  - {\bar n}_{(a)}\, {\bar n}_{(b)})\, Q_a^{-1}\, Q_b^{-1}\, V_{ra}\,
  V_{sb}\, T_{rs}\Big].
 \label{2.18}
 \eea

\medskip

Einstein's equations are not independent:\medskip

\noindent i) Four of them are the secondary super-hamiltonian and
super-momentum constraints [from Eqs.(\ref{2.3}) we have $l^A =
{1\over {1 + n}}\, \Big(1; - {\bar n}_{(a)}\, {}^3{\bar
e}^r_{(a)}\Big)$]

\medskip

\begin{eqnarray*}
 {}^4G_{ll} &=& {}^4G_{AB}\, l^A\, l^B = {1\over {(1 + n)^2}}\, [{}^4G_{\tau\tau}
 - 2\, {}^4G_{\tau r}\, n^r + {}^4G_{rs}\, n^r\, n^s] \cir\nonumber \\
 &\cir& {{8\, \pi\, G}\over {c^3}}\, T_{ll} =
 {{8\, \pi\, G}\over {c^3}}\, l^A\, l^B\, T_{AB} =\nonumber \\
 &=& {{8\, \pi\, G}\over {c^3\, (1 + n)^2}}\, \Big[T_{\tau\tau} -
 2\, {\tilde \phi}^{-1/3}\, {\bar n}_{(a)}\, Q_a^{-1}\,
 V_{ra}\, T_{\tau r} +\nonumber \\
 &+&{\tilde \phi}^{-2/3}\, {\bar n}_{(a)}\, {\bar n}_{(b)}\,
 Q_a^{-1}\, Q_b^{-1}\, V_{ra}\, V_{sb}\, T_{rs}\Big],\nonumber \\
 &&{}\nonumber \\
 0 &\cir& {}^4G_{ll} - {{8\, \pi\, G}\over {c^3}}\, T_{ll} = \sgn\,
 {\tilde \phi}^{-1}\, {{8\, \pi\, G}\over {c^3}}\, {\cal H} \approx 0,
 \end{eqnarray*}

\bea
 {}^4G_{lr} &=& {}^4G_{AB}\, l^A\, {}^3h^B_r = {}^4G_{Ar}\, l^A =
 {{\sgn}\over {1 + n}}\, [{}^4G_{\tau r} - {}^4G_{rs}\, n^s] \cir\nonumber \\
 &\cir& {{8\, \pi\, G}\over {c^3}}\, l^A\, T_{Ar} = {{8\, \pi\, G}\over
 {c^3\, (1 + n)}}\, \Big[T_{\tau r} - {\tilde \phi}^{-1/3}\, {\bar n}_{(a)}\,
 Q_a^{-1}\, V_{sa}\, T_{rs}\Big],\nonumber \\
 &&{}\nonumber \\
 0 &\cir& {}^4G_{lr} - {{8\, \pi\, G}\over {c^3}}\, T_{lr} = -
 \sgn\, {{8\, \pi\, G}\over {c^3}}\, {\tilde \phi}^{2/3}\, \sum_c\, Q_c^2\,
  V_{rc}\, V_{sc}\, {\cal H}^s \approx 0.\nonumber \\
  &&{}
  \label{2.19}
 \eea

\bigskip

\noindent ii) Four of them are not independent from the others due
to the contracted Bianchi identities (Noether identities implying
$\Rightarrow \, {}^4\nabla_A\, T^{AB}\,\, \cir\,\, 0$) generated by
the time-conservation of the super-hamiltonian and super-momentum
constraints

\medskip

\beq
 0 \equiv {}^4\nabla_A\, {}^4G^{AB} = \partial_A\, {}^4G^{AB} +
{}^4\Gamma^A_{AC}\, {}^4G^{CB} + {}^4\Gamma^B_{AC}\, {}^4G^{AC}.
 \label{2.20}
 \eeq

\medskip

At the Hamiltonian level in the York canonical basis the four
contracted Bianchi identities become the Hamilton equations for
$\tilde \phi$ and $\pi_i^{(\theta)}$, i.e. for the unknowns in the
super-Hamiltonian and super-momentum constraints.
\medskip

As a consequence of Eqs.(\ref{2.19}), Eqs.(\ref{2.20}) imply that
only two of the six components of ${}^4R_{rs}$ are independent and
give rise to genuine second order equations of motion.

\section{The Hamiltonian Expression of the 4-Riemann Tensor}

In this Section we use the material of the previous Section to find the Hamiltonian expression of the 4-Christoffel symbols and of the 4-Rienmann tensor.

\subsection{The 4-Christoffel Symbols}

After these preliminaries we can evaluate explicitly the
4-Christoffel symbols by using Eqs.(\ref{2.7})

\bea
 {}^4\Gamma^A_{BC} &=& {1\over 2}\, {}^4g^{AE}\, \Big(\partial_B\,
 {}^4g_{CE} + \partial_C\, {}^4g_{BE} - \partial_E\, {}^4g_{BC}\Big)
 =\nonumber \\
 &=&{1\over 2}\, {}^4g^{\tau A}\, \Big(\partial_B\,
 {}^4g_{\tau C} + \partial_C\, {}^4g_{\tau B} - \partial_{\tau}\, {}^4g_{BC}\Big)
 +\nonumber \\
 &+&{1\over 2}\, {}^4g^{Au}\, \Big(\partial_B\,
 {}^4g_{Cu} + \partial_C\, {}^4g_{Bu} - \partial_u\, {}^4g_{BC}\Big)
 \label{3.1}
 \eea

By using the third of Eqs.(\ref{2.15}) we get the following expressions

\begin{eqnarray*}
 {}^4\Gamma^{\tau}_{\tau\tau} &=& {1\over 2}\, \Big[{}^4g^{\tau\tau}\,
 \partial_{\tau}\, {}^4g_{\tau\tau} + {}^4g^{\tau u}\, (2\,
 \partial_{\tau}\, {}^4g_{\tau u} - \partial_u\, {}^4g_{\tau\tau})\Big]
 =\nonumber \\
 &&\nonumber\\
 &=& \frac{1}{1+n}\Big[
 \partial_\tau n+\bar{n}_{(a)}\bar{e}^r_{(a)}\partial_r n-
 \bar{n}_{(a)}\bar{n}_{(b)}\bar{e}^r_{(a)}\bar{e}^s_{(b)}\,K_{rs}
 \Big]=\nonumber\\
 &&\nonumber\\
 &=&\frac{1}{1+n}\Big\{
 \partial_\tau n+\sum_{a,r}\,\tilde{\phi}^{-1/3}Q_a^{-1}V_{ra}\bar{n}_{(a)}\partial_rn-
 \sum_{ab,rs}Q_a^{-1}Q_b^{-1}V_{ra}V_{sb}\bar{n}_{(a)}\bar{n}_{(b)}\times\nonumber\\
 &&\nonumber\\
 &&
 \times\Big[
\frac{4\pi G}{c^3}\sum_c\,Q_c^2V_{rc}V_{sc}\Big(
\pi_{\tilde{\phi}}-2\tilde{\phi}^{-1}\sum_{\bar{a}}\gamma_{\bar{a}c}\Pi_{\bar{a}}
\Big)+\sum_{c\neq d}\sigma_{(c)(d)}Q_cQ_dV_{rc}V_{sd}
 \Big]\Big\}\nonumber\\
 \end{eqnarray*}

\begin{eqnarray*}
 {}^4\Gamma^{\tau}_{\tau r} &=&{1\over 2}\, \Big[{}^4g^{\tau\tau}\, \partial_r\,
 {}^4g_{\tau\tau} + {}^4g^{\tau s}\, (\partial_{\tau}\, {}^4g_{rs} +
 \partial_r\, {}^4g_{\tau s} - \partial_s\, {}^4g_{\tau r})\Big] =\nonumber \\
 &&{}\nonumber \\
 &=& \frac{1}{1+n}\Big(\partial_r n-K_{rs}\,\bar{e}^s_{(a)}\bar{n}_{(a)}\Big)=\nonumber\\
 &&\nonumber\\
 &=&\frac{1}{1+n}\Big\{
 \partial_r n+\sum_{a,s}\,\tilde{\phi}^{1/3}Q_a^{-1}V_{sa}\times\nonumber\\
 &&\nonumber\\
 &&\times\Big[
\frac{4\pi G}{c^3}\sum_c\,Q_c^2V_{rc}V_{sc}\Big(
\pi_{\tilde{\phi}}-2\tilde{\phi}^{-1}\sum_{\bar{a}}\gamma_{\bar{a}c}\Pi_{\bar{a}}
\Big)+\sum_{c\neq d}\sigma_{(c)(d)}Q_cQ_dV_{rc}V_{sd}
 \Big]\Big\}\nonumber\\
 \end{eqnarray*}

 \begin{eqnarray*}
 {}^4\Gamma^{\tau}_{rs} &=&{1\over 2}\, \Big[{}^4g^{\tau\tau}\, (\partial_r\,
 {}^4g_{\tau s} + \partial_s\, {}^4g_{\tau r} - \partial_{\tau}\, {}^4g_{rs})
 +\nonumber \\
 &+& {}^4g^{\tau u}\, (\partial_r\, {}^4g_{su} + \partial_s\, {}^4g_{ru}
 - \partial_u\, {}^4g_{rs})\Big] =\nonumber \\
 &&{}\nonumber \\
 &=& - {1\over {1 + n}}\, {}^3K_{rs} =\nonumber \\
 &&{}\nonumber \\
 &=& - {{{\tilde \phi}^{2/3}}\over {1 + n}}\, \Big[ {{4\pi\, G}\over
 {c^3}}\, \sum_a\, Q_a^2\, V_{ra}\, V_{sa}\, \Big( \pi_{\tilde \phi}\,
 -2\, {\tilde \phi}^{-1}\, \sum_{\bar a}\, \gamma_{\bar aa}\, \Pi_{\bar a}\Big)\,
 +\nonumber \\
 &+& \sum_{ab}^{a \not= b}\, \sigma_{(a)(b)}\, Q_a\, Q_b\, V_{ra}\,
 V_{sb}\Big],
 \end{eqnarray*}

\begin{eqnarray*}
 {}^4\Gamma^u_{\tau\tau} &=&{1\over 2}\, \Big[{}^4g^{\tau u}\,
 \partial_{\tau}\, {}^4g_{\tau\tau} + {}^4g^{uv}\, (2\, \partial_{\tau}\,
 {}^4g_{\tau v} - \partial_v\, {}^4g_{\tau\tau})\Big] =\nonumber \\
 &&{}\nonumber \\
  &=&\sum_a\,\tilde{\phi}^{-1/3}Q_a^{-1}V_{ua}\,\Big\{\bar{n}_{(a)} (1+n)\partial_\tau
n-\bar{n}_{(a)}\bar{n}_{(b)}\partial_\tau\bar{n}_{(b)}+
\sum_{bc,rs}\,\frac{\bar{n}_{(a)}\bar{n}_{(b)}\bar{n}_{(c)}}{1+n}\,Q_b^{-1}Q_c^{-1}V_{rb}V_{sc}\times\nonumber\\
&&\nonumber\\
&&\times\Big[ {{4\pi\, G}\over
 {c^3}}\, \sum_d\, Q_d^2\, V_{rd}\, V_{sd}\, \Big( \pi_{\tilde \phi}\,
 -2\, {\tilde \phi}^{-1}\, \sum_{\bar a}\, \gamma_{\bar aa}\, \Pi_{\bar a}\Big)\,
 +\sum_{de}^{d \not= e}\, \sigma_{(d)(e)}\, Q_d\, Q_e\, V_{rd}\,
 V_{se}\Big]+\nonumber\\
 &&\nonumber\\
 &&+\sum_{b,r}\Big[
 (1+n)\tilde{\phi}^{-1/3}Q_b^{-1}V_{rb}\Big(\delta_{(a)(b)}-\frac{\bar{n}_{(a)}\bar{n}_{(b)}}{(1+n)^2}\Big)\partial_r n\nonumber\\
 &&\nonumber\\
 &&\qquad-\frac{\bar{n}_{(b)}}{2}\Big(
 Q_a^{-1}V_{ra}\partial_r\bar{n}_{(b)}+Q_a^{-1}V_{rb}\partial_r\bar{n}_{(a)}
 \Big)\Big]+\nonumber\\
 &&\nonumber\\
 &&+\sum_{b,s}\,\bar{n}_{(a)}\bar{n}_{(b)}\tilde{\phi}^{-1/3}Q_b^{-1}V_{sb}
 \Big(\frac{1}{3}\tilde{\phi}^{-1}\partial_s\tilde{\phi}+Q_a^{-1}\partial_sQ_a\Big)-
 \nonumber\\
 &&\nonumber\\
 &&-\sum_{b,s}\,\bar{n}_{(b)}^2\tilde{\phi}^{-1/3}Q_a^{-1}V_{sa}
 \Big(\frac{1}{3}\tilde{\phi}^{-1}\partial_s\tilde{\phi}+Q_b^{-1}\partial_sQ_b\Big)+\nonumber\\
 &&\nonumber\\
 &&+\sum_{bc,rs}\,\bar{n}_{(b)}\bar{n}_{(c)}\tilde{\phi}^{-1/3}Q_a^{-1}Q_b^{-1}Q_cV_{ra}V_{sb}
 \Big(\partial_sV_{rc}-\partial_rV_{sc}\Big) \Big\}
 \end{eqnarray*}

\begin{eqnarray*}
 {}^4\Gamma^u_{\tau r} &=&{1\over 2}\, \Big[{}^4g^{\tau u}\,
 \partial_r\, {}^4g_{\tau\tau} + {}^4g^{uv}\, (\partial_{\tau}\,
 {}^4g_{rv} + \partial_r\, {}^4g_{\tau v} - \partial_v\, {}^4g_{\tau r})\Big]
 =\nonumber \\
 &&{}\nonumber \\
 &=&{\tilde \phi}^{-1/3}\, \sum_a\, Q_a^{-1}\, V_{ua}\, \Big[-
 {\bar n}_{(a)}\, {{\partial_r\, n}\over {1 + n}} - {\tilde
 \phi}^{1/3}\, (1 + n)\, \sum_b\, (\delta_{(a)(b)} - {{{\bar n}_{(a)}\,
 {\bar n}_{(b)}}\over {(1 + n)^2}})\times\nonumber \\
 &&\Big(\sum_{c \not= b}\, Q_c\, V_{rc}\, \sigma_{(c)(b)} + {{4\pi\, G}
 \over {c^3}}\, Q_b\, V_{rb}\, (\pi_{\tilde \phi} - 2\, {\tilde \phi}^{-1}\,
 \sum_{\bar a}\, \gamma_{\bar ab}\, \Pi_{\bar a})\Big) -\nonumber \\
 &-& {1\over 2}\, \sum_{bv}\, {\bar n}_{(b)}\, \Big(Q_a^{-1}\, Q_b\,
 V_{va}\, (\partial_r\, V_{vb} - \partial_v\, V_{rb}) - Q_b^{-1}\, Q_a\,
 V_{vb}\, (\partial_r\, V_{va} - \partial_v\, V_{ra})\Big)
 +\nonumber \\
 &+& {1\over 2}\, \sum_{bcvs}\, {\bar n}_{(b)}\, Q_a^{-1}\,
 Q_b^{-1}\, Q_c^2\, V_{rc}\, V_{va}\, V_{sb}\, (\partial_v\, V_{sc} -
 \partial_s\, V_{vc}) +\sum_{b,s}\,V_{sa}V_{ra}Q_a^{-1}Q_b\partial_s\bar{n}_{(b)}\Big],\nonumber \\
 &&{}\nonumber \\
 \end{eqnarray*}

 \bea
 {}^4\Gamma^u_{rs} &=&{1\over 2}\, \Big[{}^4g^{\tau u}\, (\partial_r\,
 {}^4g_{\tau s} + \partial_s\, {}^4g_{\tau r} - \partial_{\tau}\,
 {}^4g_{rs}) +{}^4g^{uv}\, (\partial_r\, {}^4g_{sv} + \partial_s\, {}^4g_{rv} -
 \partial_v\, {}^4g_{rs})\Big] =\nonumber \\
 &&{}\nonumber \\
 &=&{}^3\Gamma^u_{rs} + {{{\bar n}_{(a)}}\over {1 + n}}\,
 {}^3{\bar e}^u_{(a)}\, {}^3K_{rs} =\nonumber \\
 &&{}\nonumber \\
 &=&{}^3\Gamma^u_{rs} + {\tilde \phi}^{1/3}\, \sum_c\, Q_c^{-1}\,
 V_{uc}\, {{{\bar n}_{(c)}}\over {1 + n}}\, \Big[ {{4\pi\, G}\over
 {c^3}}\, \sum_a\, Q_a^2\, V_{ra}\, V_{sa}\, \Big( \pi_{\tilde \phi}\,
 -2\, {\tilde \phi}^{-1}\, \sum_{\bar a}\, \gamma_{\bar aa}\, \Pi_{\bar a}\Big)\,
 +\nonumber \\
 &+& \sum_{ab}^{a \not= b}\, \sigma_{(a)(b)}\, Q_a\, Q_b\, V_{ra}\,
 V_{sb}\Big].
 \label{3.2}
 \eea

${}^4\Gamma^{\tau}_{\tau\tau}$ and ${}^4\Gamma^u_{\tau\tau}$ depend on the arbitrary velocities $\partial_{\tau}\, n$
and $\partial_{\tau}\, {\bar n}_{(a)}$ of Eqs.(\ref{2.14}), which are gauge-dependent quantities determined only
after a gauge fixing.

\subsection{The 4-Riemann and 4-Ricci Tensors}

See Refs. \cite{19,20} for the symmetries and the Bianchi identities satisfied by the 4-Riemann tensor and for the derivation of the Gauss, Codazzi-Mainardi and Ricci equations. We only rewrite them in our 3+1 splitting of the space-time in radar 4-coordinates.\medskip

The radar 4-Riemann tensor with 4-scalar components, i.e. the
4-Riemann tensor in adapted radar coordinates is (to simplify some formulas we use the
notation $n_r$ for $n_r = {\bar n}_{(a)}\, {}^3{\bar e}_{(a)r} = {\tilde \phi}^{1/3}\, {\bar n}_{(a)}\, Q_a\, V_{ra}$)

\medskip

\begin{eqnarray*}
 {}^4R^A{}_{BCD} &=&  \partial_C\, {}^4\Gamma^A_{BD} - \partial_D\,
{}^4\Gamma^A_{BC} + {}^4\Gamma^E_{BD}\, {}^4\Gamma^A_{CE} -
{}^4\Gamma^E_{BC}\, {}^4\Gamma^A_{DE},\nonumber \\
 &&{}\nonumber \\
 &&{}\nonumber \\
 {}^4R_{ABCD} &=& {}^4g_{AE}\, {}^4R^E{}_{BCD} = {}^4g_{AE}\, \Big(\partial_C\,
{}^4\Gamma^E_{BD} - \partial_D\, {}^4\Gamma^E_{BC} +
{}^4\Gamma^F_{BD}\, {}^4\Gamma^E_{CF} - {}^4\Gamma^F_{BC}\,
{}^4\Gamma^E_{DF}\Big) =\nonumber \\
 &=& - {1\over 2}\, \Big(\partial_A\, \partial_C\, {}^4g_{BD} + \partial_B\,
\partial_D\, {}^4g_{AC} - \partial_A\, \partial_D\, {}^4g_{BC} -
\partial_B\, \partial_C\, {}^4g_{AD}\Big) +\nonumber \\
 &+& {}^4g_{EF}\, \Big({}^4\Gamma^E_{AD}\,
{}^4\Gamma^F_{BC} - {}^4\Gamma^E_{AC}\, {}^4\Gamma^F_{BD}\Big) =\nonumber \\
 &=& - {1\over 2}\, \Big(\partial_A\, \partial_C\, {}^4g_{BD} + \partial_B\,
\partial_D\, {}^4g_{AC} - \partial_A\, \partial_D\, {}^4g_{BC} -
\partial_B\, \partial_C\, {}^4g_{AD}\Big) +\nonumber \\
 &+& \sgn\, \Big([(1 + n)^2 - {\bar n}_{(a)}\, {\bar n}_{(a)}]\,
\Big[{}^4\Gamma^{\tau}_{AD}\, {}^4\Gamma^{\tau}_{BC} -
{}^4\Gamma^{\tau}_{AC}\, {}^4\Gamma^{\tau}_{BD}\Big] -
  {}^3g_{uv}\, \Big[{}^4\Gamma^u_{AD}\, {}^4\Gamma^v_{BC} - {}^4\Gamma^u_{AC}\,
{}^4\Gamma^v_{BD}\Big] -\nonumber \\
 &-& n_u\, \Big[{}^4\Gamma^{\tau}_{AD}\, {}^4\Gamma^u_{BC} + {}^4\Gamma^u_{AD}\,
{}^4\Gamma^{\tau}_{BC} - {}^4\Gamma^{\tau}_{AC}\, {}^4\Gamma^u_{BD}
- {}^4\Gamma^u_{AC}\, {}^4\Gamma^{\tau}_{BD}\Big]\Big) =
 \end{eqnarray*}

\begin{eqnarray*}
 &=& - {1\over 2}\, \Big(\partial_A\, \partial_C\, {}^4g_{BD} + \partial_B\,
 \partial_D\, {}^4g_{AC} - \partial_A\, \partial_D\, {}^4g_{BC} -
 \partial_B\, \partial_C\, {}^4g_{AD}\Big) +\nonumber \\
 &+& \sgn\, \Big({}^4\Gamma^{\tau}_{AD}\,
 \Big[((1 + n)^2 - {\bar n}_{(a)}\, {\bar n}_{(a)})\,
 {}^4\Gamma^{\tau}_{BC} - n_u\, {}^4\Gamma^u_{BC}\Big]
 - \nonumber \\
 &-& {}^4\Gamma^{\tau}_{AC}\, \Big[((1 + n)^2 - {\bar n}_{(a)}\,
 {\bar n}_{(a)})\, {}^4\Gamma^{\tau}_{BD} - n_u\,
 {}^4\Gamma^u_{BD}\Big] -\nonumber \\
 &-& {}^4\Gamma^u_{AD}\, ({}^3g_{uv}\,
 {}^4\Gamma^v_{BC} + n_u\, {}^4\Gamma^{\tau}_{BC}) +
 {}^4\Gamma^u_{AC}\, ({}^3g_{uv}\, {}^4\Gamma^v_{BD} + n_u\,
 {}^4\Gamma^{\tau}_{BD})\Big),
 \end{eqnarray*}

\bea
 &&l^A\, {}^4R_{ABCD} = {1\over {1 + n}}\, \Big({}^4R_{\tau BCD}
 - n^s\, {}^4R_{sBCD}\Big),\nonumber \\
 &&{}\nonumber \\
 &&l^A\, {}^4R_{Aruv} = \sgn\, \Big({}^3K_{ru|v} - {}^3K_{rv|u}\Big),
 \qquad (CODAZZI-MAINARDI EQUATION),\nonumber \\
 &&{}
 \label{3.3}
 \eea

\bigskip

The radar 4-Ricci tensor ($S_{AB}$ is the trace-free Ricci tensor) and the
4-curvature scalar ($\Lambda = {}^4R/24$ is the Newman-Penrose notation \cite{17})
are

\begin{eqnarray*}
 {}^4R_{AB} &=&   {}^4R_{BA} = {}^4g^{EF}\, {}^4R_{EAFB} =
 \partial_C\, {}^4\Gamma^C_{AB} - \partial_B\, {}^4\Gamma^C_{AC} +
 {}^4\Gamma^C_{CD}\, {}^4\Gamma^D_{AB} - {}^4\Gamma^C_{BD}\,
 {}^4\Gamma^D_{CA},\nonumber \\
 &&{}\nonumber \\
 {}^4R &=&  {}^4g^{AB}\, {}^4R_{AB} = 24\, \Lambda,\qquad
 {}^4S_{AB} = {}^4R_{AB} - {1\over 4}\, {}^4g_{AB}\, {}^4R,
 \end{eqnarray*}

 \begin{eqnarray*}
  {}^4R_{\tau\tau} &=& {}^4g^{EF}\, {}^4R_{E\tau F\tau} = {}^4g^{rs}\, {}^4R_{\tau
 r\tau s} =  \nonumber \\
 &=& - \sgn\, {\tilde \phi}^{-2/3}\, Q_a^{-1}\, Q_b^{-1}\, V_{ra}\,
 V_{sb}\, (\delta_{ab} - {{{\bar n}_{(a)}\, {\bar n}_{(b)}}\over {(1 + n)^2}}) {}^4R_{\tau r \tau s} =
 E_{\tau\tau} + {{8\pi G}\over {c^3}}\, {\hat T}_{\tau\tau},\nonumber \\
  {}^4R_{\tau r} &=& {}^4g^{EF}\, {}^4R_{E\tau Fr} = - {}^4g^{\tau s}\, {}^4R_{\tau
 r\tau s} - {}^4g^{uv}\, {}^4R_{\tau urv} =\nonumber \\
 &=&\sgn\, {\tilde \phi}^{-1/3}\, Q_a^{-1}\, V_{ua}\, \Big[{{{\bar
 n}_{(a)}}\over {(1 + n)^2}}\, {}^4R_{\tau r\tau u} + {\tilde
 \phi}^{-1/3}\, Q_b^{-1}\, V_{vb}\, (\delta_{ab} - {{{\bar n}_{(a)}\,
 {\bar n}_{(b)}}\over {(1 + n)^2}})\, {}^4R_{\tau urv}\Big]
 =\nonumber \\
 &=& E_{\tau r} + {{8\pi G}\over {c^3}}\, {\hat T}_{\tau r},\nonumber \\
  {}^4R_{rs} &=& {}^4g^{EF}\, {}^4R_{ErFs} =\nonumber \\
 &=& \sgn\, \Big[{1\over {(1 + n)^2}}\, {}^4R_{\tau r\tau s} - {\tilde \phi}^{-1/3}\,
 {{Q_a^{-1}\, V_{ua}\, {\bar n}_{(a)}}\over {(1 + n)^2}}\,
 ({}^4R_{\tau rus} + {}^4R_{\tau sur}) -\nonumber \\
 &-& {\tilde \phi}^{-2/3}\, Q_a^{-1}\, Q_b^{-1}\, V_{ua}\, V_{vb}\,
  (\delta_{ab} - {{{\bar n}_{(a)}\, {\bar n}_{(b)}}\over {(1 +
  n)^2}})\, {}^4R_{rusv}\Big] = E_{rs} + {{8\pi G}\over {c^3}}\,
  {\hat T}_{rs},\nonumber \\
 \end{eqnarray*}

 \bea
  {}^4R &=& {}^4g^{\tau\tau}\, {}^4R_{\tau\tau} + 2\, {}^4g^{\tau r}\, {}^4R_{\tau r} +
 {}^4g^{rs}\, {}^4R_{rs} =\nonumber \\
 &=& \sgn\, \Big[{1\over {(1 + n)^2}}\, {}^4R_{\tau\tau} - 2\,
 {\tilde \phi}^{-1/3}\, {{Q_a^{-1}\, V_{ra}\, {\bar n}_{(a)}}\over {(1 + n)^2}}\,
 {}^4R_{\tau r} -\nonumber \\
 &-& {\tilde \phi}^{-2/3}\, Q_a^{-1}\, Q_b^{-1}\, V_{ra}\, V_{sb}\,
  (\delta_{ab} - {{{\bar n}_{(a)}\, {\bar n}_{(b)}}\over {(1 +
  n)^2}})\, {}^4R_{rs}\Big] = {}^4g^{AB}\, E_{AB} - {{8\pi G}\over
  {c^3}}\, T.
 \label{3.4}
 \eea

\bigskip

The components of the radar 4-Riemann tensor are

\begin{eqnarray*}
  {}^4R_{rsuv} &=& - {{\sgn}\over 2}\, \Big(\partial_r\, \partial_v\, {}^3g_{su} +
\partial_s\, \partial_u\, {}^3g_{rv} - \partial_r\, \partial_u\, {}^3g_{sv} -
\partial_s\, \partial_v\, {}^3g_{ru}\Big) -\nonumber \\
 &-& \sgn\, \Big[((1 + n)^2 - {\bar n}_{(a)}\, {\bar n}_{(a)})\,
 \Big({}^4\Gamma^{\tau}_{ru}\, {}^4\Gamma^{\tau}_{sv} -
 {}^4\Gamma^{\tau}_{rv}\, {}^4\Gamma^{\tau}_{su}\Big) +
 {}^3g_{hk}\, \Big({}^4\Gamma^h_{rv}\, {}^4\Gamma^k_{su} -
 {}^4\Gamma^h_{ru}\, {}^4\Gamma^k_{sv}\Big) +\nonumber \\
 &+& n_h\, \Big({}^4\Gamma^{\tau}_{rv}\, {}^4\Gamma^h_{su} +
 {}^4\Gamma^{\tau}_{su}\, {}^4\Gamma^h_{rv} -
 {}^4\Gamma^{\tau}_{ru}\, {}^4\Gamma^h_{sv} -
 {}^4\Gamma^{\tau}_{sv}\, {}^4\Gamma^h_{ru}\Big)\Big] =\nonumber \\
 &=& - \sgn\, \Big({}^3R_{rsuv} + {}^3K_{rv}\, {}^3K_{su} -
 {}^3K_{ru}\, {}^3K_{sv}\Big)\, =\qquad (GAUSS\,\, EQUATION)\nonumber \\
 &{\buildrel {def}\over =}&\, {\bar W}_{rsuv} = - {\bar W}_{sruv} = - {\bar W}_{rsvu} =
 {\bar W}_{uvrs},\nonumber \\
 &&{}
 \end{eqnarray*}

\begin{eqnarray*}
  {}^4R_{\tau ruv} &=& - {{\sgn}\over 2}\, \Big(\partial_r\,
 \partial_u\, n_v + \partial_v\, \partial_{\tau}\, {}^3g_{ru} -
 \partial_r\, \partial_v\, n_u - \partial_u\, \partial_{\tau}\,
 {}^3g_{rv}\Big) -\nonumber \\
 &-& \sgn\, \Big[((1 + n)^2 - {\bar n}_{(a)}\,
 {\bar n}_{(a)})\, \Big({}^4\Gamma^{\tau}_{\tau u}\,
 {}^4\Gamma^{\tau}_{rv} - {}^4\Gamma^{\tau}_{\tau v}\,
 {}^4\Gamma^{\tau}_{ru} \Big) + {}^3g_{hk}\, \Big(
 {}^4\Gamma^k_{ru}\, {}^4\Gamma^h_{\tau v} - {}^4\Gamma^k_{rv}\,
 {}^4\Gamma^h_{\tau u} \Big) +\nonumber \\
 &+& n_h\, \Big({}^4\Gamma^{\tau}_{ru}\, {}^4\Gamma^h_{\tau v}
 + {}^4\Gamma^{\tau}_{\tau v}\, {}^4\Gamma^h_{ru} -
 {}^4\Gamma^{\tau}_{rv}\, {}^4\Gamma^h_{\tau u} -
 {}^4\Gamma^{\tau}_{\tau u}\, {}^4\Gamma^h_{rv} \Big)
 \Big] =\nonumber \\
 &=& (1 + n)\, l^A\, {}^4R_{Aruv} + n^s\, {}^4R_{sruv} =\nonumber \\
 &=& \sgn\, \Big[(1 + n)\, \Big({}^3K_{ru|v} - {}^3K_{rv|u}\Big) +
 {\tilde \phi}^{-1/3}\, Q_a^{-1}\, V_{sa}\, {\bar n}_{(a)}\,
  \Big({}^3R_{rsuv} + {}^3K_{rv}\, {}^3K_{su} -
 {}^3K_{ru}\, {}^3K_{sv}\Big)\Big] =\nonumber \\
 &{\buildrel {def}\over =}&\, {\bar W}_{\tau ruv} = - {\bar W}_{\tau rvu},
 \nonumber \\
 &&{}
\end{eqnarray*}

\bea
 {}^4R_{\tau r\tau s} &=& - {{\sgn}\over 2}\, \Big(\partial_r\,
 \partial_s\, [(1 + n)^2 - {\bar n}_{(a)}\, {\bar n}_{(a)}] -
 \partial^2_{\tau}\, {}^3g_{rs} +\partial_r\, \partial_{\tau}\,
 n_s + \partial_s\, \partial_{\tau}\, n_r\Big) -\nonumber \\
 &-& \sgn\, \Big[((1 + n)^2 - {\bar n}_{(a)}\, {\bar n}_{(a)})\,
\Big({}^4\Gamma^{\tau}_{\tau\tau}\, {}^4\Gamma^{\tau}_{rs} -
{}^4\Gamma^{\tau}_{\tau r}\, {}^4\Gamma^{\tau}_{\tau s}\Big) +
{}^3g_{hk}\, \Big({}^4\Gamma^h_{\tau r}\, {}^4\Gamma^k_{\tau s} -
{}^4\Gamma^h_{\tau\tau}\, {}^4\Gamma^k_{rs}\Big) +\nonumber \\
 &+& n_h\, \Big({}^4\Gamma^{\tau}_{\tau r}\, {}^4\Gamma^h_{\tau s}
 + {}^4\Gamma^{\tau}_{\tau s}\, {}^4\Gamma^h_{\tau r} -
{}^4\Gamma^{\tau}_{\tau\tau}\, {}^4\Gamma^h_{rs} -
{}^4\Gamma^{\tau}_{rs}\, {}^4\Gamma^h_{\tau\tau} \Big) \Big].
 \label{3.5}
 \eea

To get the Hamiltonian expression of ${}^4R_{rsuv}$ and of ${}^4R_{\tau ruv}$
we must use Eq.(\ref{2.11}) for ${}^3K_{rs}$, Eqs. (\ref{a1}) for
${}^3R_{rsuv}$, Eq.(\ref{2.12}) for ${}^3\Gamma^u_{rs}$ and Eq.(\ref{2.13}) for
${}^3K_{rs|u} = \partial_u\, {}^3K_{rs} -
{}^3\Gamma^v_{ru}\, {}^3K_{vs} - {}^3\Gamma^v_{su}\, {}^3K_{rv}$. These quantities do not depend
on the first two arbitrary velocities of Eq.(\ref{2.15}).
The first of Eqs.(\ref{3.5}) is the Gauss equation, while the second one is a combination of the
Codazzi-Mainardi equation with the Gauss one. As a consequence ${\bar
W}_{rsuv}$ and ${\bar W}_{\tau ruv}$ are well defined Hamiltonian functions.

\medskip

The third equation is the analogue of the Ricci equation in radar
4-coordinates: it cannot be expressed in terms of the canonical
variables without using the equations of motion (the dynamical
second half of Hamilton equations) because it contains
$\partial^2_{\tau}\,\,\, {}^3g_{rs}$ besides the $\tau$-derivatives
of the lapse and shift functions (both explicitly in the first line
and implicitly inside the 4-Christoffel symbols
${}^4\Gamma^{\tau}_{\tau\tau}$ and ${}^4\Gamma^u_{\tau\tau}$ given in Eqs.(\ref{3.2})) given in the first two of Eqs.(\ref{2.15}).

\medskip

However Eq.(\ref{3.4}) implies

\begin{eqnarray*}
 {}^4R_{\tau r\tau s} &=& \Big({}^4R_{\tau rus} + {}^4R_{\tau sur}\Big)\, n^u
 + {}^4R_{rusv}\, \Big((1 + n)^2\,\, {}^3g^{uv} - n^u\, n^v\Big) +
 \sgn\, (1 + n)^2\,\, {}^4R_{rs} =\nonumber \\
 &&{}\nonumber \\
 &=&\sgn\, \Big[{\tilde \phi}^{-1/3}\,  (1 + n)\, \Big(2\,\, {}^3K_{rs|u} - {}^3K_{ur|s} -
 {}^3K_{us|r}\Big)\, Q_a^{-1}\, V_{ua}\, {\bar n}_{(a)} +\nonumber \\
 &+& (1 + n)^2\, \Big({}^3R_{rs} + {\tilde \phi}^{-2/3}\, Q_a^{-2}\, V_{ua}\, V_{va}\,
 {}^3K_{ru}\,\,  {}^3K_{vs} - {}^3K\,\, {}^3K_{rs}\Big) +\nonumber \\
 &+& {\tilde \phi}^{-2/3}\, \Big({}^3R_{ruvs} + {}^3K_{rs}\,\, {}^3K_{uv} - {}^3K_{ru}\,\,
 {}^3K_{sv}\Big)\, Q_a^{-1}\, Q_b^{-1}\, V_{ua}\, V_{vb}\, {\bar n}_{(a)}\, {\bar n}_{(b)}\Big]
 + \sgn\, (1 + n)^2\,\, {}^4R_{rs} =\nonumber \\
 &{\buildrel {def}\over =}&\,\, {\bar W}_{\tau r\tau s} + \sgn\, (1 + n)^2\,\,
 {}^4R_{rs} =\nonumber \\
 &=&{\bar W}_{\tau r\tau s} + \sgn\, (1 + n)^2\,\, \Big[E_{rs} +
 {{8\pi G}\over {c^3}}\, {\hat T}_{rs}\Big],
 \end{eqnarray*}

 \bea
  &&\Downarrow\nonumber \\
  &&{}\nonumber \\
 {\bar W}_{\tau r\tau s} &=& ({\bar W}_{\tau rus} + {\bar W}_{\tau sur})\, n^u +
 {\bar W}_{rusv}\, ((1 + n)^2\,\, {}^3g^{uv} - n^u\, n^v) =\nonumber \\
 &=&\sgn\, \Big[{\tilde \phi}^{-1/3}\, (1 + n)\, Q_a^{-1}\, V_{ua}\, {\bar n}_{(a)}\,
 \Big(2\, {}^3K_{rs|u} - {}^3K_{ur|s} - {}^3K_{us|r}\Big) +\nonumber \\
 &+&   (1 + n)^2\, \Big({}^3R_{rs} + {\tilde \phi}^{-2/3}\, Q_a^{-2}\, V_{ua}\, V_{va}\,
 {}^3K_{rv}\, {}^3K_{su} - {}^3K\, {}^3K_{rs}\Big) +\nonumber \\
 &+&  {\tilde \phi}^{-2/3}\, Q_a^{-1}\, Q_b^{-1}\, V_{ua}\, V_{vb}\, {\bar n}_{(a)}\,
 {\bar n}_{(b)}\, \Big({}^3R_{ruvs} + {}^3K_{rs}\, {}^3K_{uv} -
 {}^3K_{ru}\, {}^3K_{sv}\Big)\Big],\nonumber \\
  &&{}\nonumber \\
  &&{}\nonumber \\
  {}^4R_{\tau\tau} &=& - \sgn\, \Big({}^3g^{rs} - {{n^r\, n^s}\over {(1 + n)^2}}\Big)\,
  {\bar W}_{\tau r\tau s} - \Big((1 + n)^2\,\, {}^3g^{rs} - n^r\, n^s\Big)\,
  {}^4R_{rs} =\nonumber \\
  &=& E_{\tau\tau} + {{8\pi G}\over {c^3}}\, {\hat T}_{\tau\tau},\nonumber \\
  {}^4R_{\tau r} &=& \sgn\, \Big[{\bar W}_{\tau r\tau s}\, {{n^s}\over {(1 + n)^2}} + \Big({}^3g^{uv}
  - {{n^u\, n^v}\over {(1 + n)^2}}\Big)\, {\bar W}_{\tau urv}\Big] + {}^4R_{rs}\,
  n^s =\nonumber \\
  &=& E_{\tau r} + {{8\pi G}\over {c^3}}\, {\hat T}_{\tau r},\nonumber \\
  {}^4R &=& - {1\over {(1 + n)^2}}\, \Big[\Big({}^3g^{uv} + {{n^u\, n^v}\over {(1 + n)^2}}\Big)\,
  {\bar W}_{\tau u\tau v} + 2 n^r\, \Big({}^3g^{uv} - {{n^u\, n^v}\over {(1 + n)^2}}\Big)\,
  {\bar W}_{\tau urv}\Big] - 2 \sgn\, {}^3g^{rs}\,  {}^4R_{rs} =\nonumber \\
  &=& {}^4g^{AB}\, E_{AB} + {{4\pi G}\over {c^3}}\, T.\nonumber \\
  &&{}
 \label{3.6}
 \eea

\bigskip

As a consequence the 4-scalar radar quantities  ${\bar W}_{\tau r\tau s}\, {\buildrel
{def}\over =}\, {}^4R_{\tau r\tau s} - \sgn\, (1 + n)^2\,
{}^4R_{rs}$, ${\bar W}_{\tau ruv}\, {\buildrel {def}\over =}\,
{}^4R_{\tau ruv}$ and ${\bar W}_{rsuv}\, {\buildrel {def}\over =}\,
{}^4R_{rsuv}$ are expressible in terms of the canonical variables.
\medskip

While Eqs.(\ref{3.5}) imply that ${\bar W}_{\tau ruv}$ and ${\bar
W}_{rsuv}$ have the same symmetries of ${}^4R_{\tau ruv}$ and
${}^4R_{rsuv}$, respectively, ${\bar W}_{\tau r\tau s}$ has not the
symmetries of ${}^4R_{\tau r\tau s}$ except ${\bar W}_{\tau r\tau s}
= {\bar W}_{\tau s\tau r}$.

\medskip

Therefore we can introduce a Hamiltonian radar tensor ${\bar {\cal W}}_{ABCD}$

\bea
 {\bar {\cal W}}_{ABCD} &=&{}^4R_{ABCD} - \delta^{\tau}_A\, \delta^r_B\,
 \delta^{\tau}_C\, \delta^s_D\, \epsilon\, (1 + n)^2\, E_{rs} \cir {}^4R_{ABCD},\nonumber \\
 {}&&\nonumber \\
&&{\bar {\cal W}}_{rsuv} = {\bar W}_{rsuv},\qquad {\bar {\cal W}}_{\tau ruv} = {\bar W}_{\tau ruv},\nonumber \\
&&{\bar {\cal W}}_{\tau r\tau s} = {\bar W}_{\tau r\tau s} + \epsilon\, {{8 \pi G}\over {c^3}}\, (1 + n)^2\, {\hat T}_{rs},
\label{3.7}
\eea

\noindent which becomes the radar 4-Riemann tensor "on-shell", namely by using Einstein's equations.

\medskip

In conclusion the 20 independent components of the radar 4-Riemann tensor
${}^4R_{ABCD}$ are expressible in terms of the canonical quantities
${\bar W}_{ABCD}$ and of ${}^4R_{rs}$, which has the following
expression due to Eq.(\ref{2.18})

\bea
 {}^4R_{rs} &=& E_{rs} + {{8\pi G}\over {c^3}}\, {\hat T}_{rs} =
 E_{rs} + {{8\, \pi\, G}\over {c^3}}\,
  \Big[T_{rs} - {{\sgn}\over 2}\, {\tilde \phi}^{2/3}\, \sum_c\, Q_c^2\,
  V_{rc}\, V_{sc}\, T\Big] \cir   \nonumber \\
  &&{\cir}  {{8\, \pi\, G}\over {c^3}}\,
  \Big[T_{rs} - {{\sgn}\over 2}\, {\tilde \phi}^{2/3}\, \sum_c\, Q_c^2\,
  V_{rc}\, V_{sc}\, T\Big].
  \label{3.8}
  \eea

\medskip

Einstein's equations (\ref{2.19}) imply:

 \bea
  && {1 \over 2}\,({}^3g^{rs} + {{n^r\, n^s}\over {(1 + n)^2}})\,
 {\bar W}_{\tau r\tau s} + n^r\, {}^3g^{uv}\, {\bar W}_{\tau urv}
 \approx\nonumber \\
 &&\qquad \approx - \sgn\, {{8\, \pi\, G}\over {c^3}}\, \Big[T_{\tau\tau} -
 2\, n^r\, T_{\tau r} + n^r\, n^s\, T_{rs}\Big],\nonumber \\
 &&{}\nonumber \\
 &&{\bar W}_{\tau r\tau s}\, {{n^s}\over {(1 + n)^2}} +
 ({}^3g^{uv} - {{n^u\, n^v}\over {(1 + n)^2}})\, {\bar W}_{\tau urv} \approx
 \nonumber \\
 &&\qquad \approx \sgn\, {{8\, \pi\, G}\over
 {c^3\, (1 + n)}}\, \Big[T_{\tau r} - n^s\, T_{rs}\Big].
 \label{3.9}
 \eea
 \medskip

As a consequence 4 of the 6 components of ${\bar W}_{\tau r\tau s}$
are weakly determined from the other $\bar W$'s and from the matter
energy-momentum tensor.

\section{Null Tetrads and the 4-Ricci Scalars of the Newman-Penrose Approach}

In this Section we introduce a set of Hamiltonian null tetrads suggested by the
framework of ADM tetrad gravity. They are the natural tools to get a Hamiltonian formulation
of the Newman-Penrose formalism \cite{17}. Here we give the Hamiltonian expression of the 4-Ricci scalars as
sums of terms in the Hamiltonian energy-momentum tensor of matter plus terms vanishing with
Einstein's equations.

\subsection{Null Tetrads.}

Given the 3+1 splitting of the space-time and the reference adapted
tetrads (\ref{2.4}), we see that the canonical formalism
automatically identifies the following unit time-like vector $l^A$
and unit space-like vector ${\cal {\bar  N}}^A$

\bigskip

\bea
 l^A &=& {}^4{\buildrel \circ \over {\bar E}}^A_{(o)} =
 {1\over {1 + n}}\, \Big( 1;\,\, - {\bar n}_{(a)}\, {}^3{\bar e}^r_{(a)}
\Big), \qquad {}^4g_{AB}\, l^A\, l^B = \sgn,\nonumber \\
 &&{}\nonumber \\
 {\cal {\bar N}}^A &=& {\hat {\bar n}}_{(a)}\, {}^4{\buildrel
 \circ \over {\bar E}}^A_{(a)} =
 \Big( 0;\,\, {\hat {\bar n}}_{(a)}\, {}^3{\bar e}^r_{(a)}\Big), \qquad
{}^4g_{AB}\, {\cal {\bar N}}^A\, {\cal {\bar N}}^B = - \sgn,\nonumber  \\
 &&{}\nonumber \\
 &&\qquad\qquad {\hat {\bar n}}_{(a)} = {{{\bar n}_{(a)}}\over {\sqrt{\sum_c\,
{\bar n}^2_{(c)}}}},\qquad \sum_a\, {\hat {\bar n}}_{(a)}^2 = 1,\nonumber \\
 &&{}\nonumber \\
 l_A &=& \sgn\, {}^4{\buildrel \circ \over {\bar E}}^{(o)}_A =
 \sgn\, (1 + n)\, \Big(1;\,\, 0\Big),\nonumber \\
 &&{}\nonumber \\
 {\cal {\bar N}}_A &=& - \sgn\, \Big(\sqrt{\sum_c\,
{\bar n}^2_{(c)}};\,\, {\hat {\bar n}}_{(a)}\, {}^3{\bar
e}_{(a)r}\Big).
 \label{4.1}
 \eea

\medskip

Given the shift unit 3-vector ${\hat {\bar n}}_{(a)}$ in the
Euclidean 3-space with flat indices $(a)$, we can define the
following basis of three unit vectors ($R$ is an arbitrary rotation
matrix identifying the "3" axis)\medskip

\begin{eqnarray*}
  \delta_{(a)(b)} &=& {\hat {\bar n}}_{(a)}\, {\hat {\bar n}}_ {(b)}
+ {\hat {\bar \epsilon}}_{(1)(a)}\, {\hat {\bar \epsilon}}_{(1)(b)}
+ {\hat {\bar  \epsilon}}_{(1)(a)}\,
{\hat {\bar \epsilon}}_{(1)(b)},\nonumber \\
 &&{}\nonumber \\
 && \sum_a\, {\hat {\bar \epsilon}}^2_{(1)(a)} =
 \sum_a\, {\hat {\bar \epsilon}}^2_{(2)(a)}
 = 1,   \qquad \sum_{(a)} {\hat {\bar n}}_{(a)}\,
 {\hat {\bar \epsilon}}_{(1,2)(a)} = 0,
\qquad \sum_a\, {\hat {\bar \epsilon}}_{(1)(a)}\, {\hat {\bar
\epsilon}}_{(2)(a)} = 0,
 \end{eqnarray*}

\begin{eqnarray*}
 {\hat {\bar n}}_{(a)} &=& (\,\,R\, \left( \begin{array}{l} 0\\ 0\\ 1
\end{array} \right)\,\, )_{(a)},\nonumber \\
 &&{}\nonumber \\
 {\hat {\bar \epsilon}}_{(1)(a)} &=& (\,\, R\, \left( \begin{array}{l} 1\\ 0\\ 0
\end{array} \right)\,\, )_{(a)} = {1\over {\sqrt{1 - {\hat
{\bar n}}^2_{(3)}}}}\, \left( \begin{array}{l} {\hat {\bar
n}}_{(1)}\, {\hat {\bar n}}_{(3)} \\ {\hat {\bar n}}_{(2)}\, {\hat
{\bar n}}_{(3)} \\ - [1 - {\hat
{\bar n}}^2_{(3)}]\end{array} \right) =\nonumber \\
 &=&{1\over {\sqrt{({\bar n}^2_{(1)} + {\bar n}^2_{(2)})\,
\sum_c\, {\bar n}^2_{(c)}}}}\, \left( \begin{array}{l} {\bar
n}_{(1)}\, {\bar n}_{(3)} \\ {\bar n}_{(2)}\,
{\bar n}_{(3)} \\ - {\bar n}^2_{(1)} -
{\bar n}^2_{(2)}\end{array} \right),\nonumber \\
   \end{eqnarray*}

\bea
 {\hat {\bar \epsilon}}_{(2)(a)} &=& (\,\, R\, \left( \begin{array}{l} 0\\ 1\\ 0
\end{array} \right)\,\, )_{(a)} = {1\over {\sqrt{1 - {\hat
{\bar n}}^2_{(3)}}}}\, \left( \begin{array}{l} - {\hat {\bar n}}_{(2)} \\
{\hat {\bar n}}_{(1)} \\ 0\end{array} \right) = {1\over {\sqrt{
{\bar n}^2_{(1)} + {\bar n}^2_{(2)}}}}\, \left( \begin{array}{l} - {\bar n}_{(2)} \\
{\bar n}_{(1)} \\ 0\end{array} \right),\nonumber \\
 {}&&
 \label{4.2}
 \eea

\noindent and then we can replace ${\hat {\bar \epsilon}}_{(1)(a)}$
and ${\hat {\bar \epsilon}}_{(2)(a)}$ with a circular basis

\bea
 {\hat {\bar \epsilon}}_{(\pm )(a)} &=& {1\over {\sqrt{2}}}\, ({\hat
{\bar \epsilon}}_{(1)(a)} \pm i\, {\hat {\bar \epsilon}}_{(2)(a)}) =
{1\over {\sqrt{2}\, \sqrt{1 - {\hat {\bar n}}^2_{(3)}}}}\, \left(
\begin{array}{l} {\hat {\bar n}}_{(1)}\, {\hat {\bar n}}_{(3)} \mp
i\, {\hat {\bar n}}_{(2)} \\ {\hat {\bar n}}_{(2)}\, {\hat {\bar
n}}_{(3)} \pm i\, {\hat {\bar n}}_{(1)}
\\ - [1 - {\hat {\bar n}}^2_{(3)}] \end{array} \right),\nonumber \\
 &&{}\nonumber \\
 &&\sum_a\, {\hat {\bar n}}_{(a)}\, {\hat {\bar
 \epsilon}}_{(\pm)(a)} = 0,\qquad
 \sum_a\, {\hat {\bar \epsilon}}_{(\pm )(a)}\, {\hat {\bar \epsilon}}_{(\pm
)(a)} = 0,\qquad \sum_a\, {\hat {\bar \epsilon}}_{(+)(a)}\, {\hat
{\bar \epsilon}}_{(-)(a)} = 1,\nonumber \\
 \delta_{(a)(b)} &=&  {\hat {\bar n}}_{(a)}\, {\hat {\bar n}}_
{(b)} + {\hat {\bar \epsilon}}_{(+)(a)}\, {\hat {\bar
\epsilon}}_{(-)(b)} + {\hat {\bar  \epsilon}}_{(-)(a)}\, {\hat {\bar
\epsilon}}_{(+)(b)}.
 \label{4.3}
 \eea

\bigskip

Therefore, in each point of every instantaneous 3-space
$\Sigma_{\tau}$ there is a spatial direction identified by the shift
functions ${\bar n}_{(a)}$, i.e. by the gauge variables describing
gravitomagnetism \footnote{If there exist gauges (in particular
coordinate systems) in which the vector field ${\cal {\bar
N}}^A(\tau ,\vec \sigma )\, \partial_A = ({\hat {\bar n}}_{(a)}\,
{}^3{\bar e}^r_{(a)})(\tau ,\vec \sigma )\,
\partial_r$ is surface-forming (i.e. it has zero vorticity), then
in these gauges each $\Sigma_{\tau}$ would be foliated with
2-surfaces having ${\cal {\bar N}}(\tau ,\vec \sigma )$ as unit
3-normal and the 3+1 splitting would become a (2+1) + 1 = 2 + 2
splitting \cite{25,26}. Since the surface-forming condition is
$d[{\cal {\bar N}}_A(\tau ,\vec \sigma )\, d\sigma^A] = 0$ (this
1-form must be closed), we have that these gauges exist if there are
solutions to the equations ${\hat {\bar n}}_{(a)}\, {}^3{\bar
e}_{(a)r} = \partial_r\, f$ and $ \sqrt{\sum_a\, {\bar n}^2_{(a)}} =
\partial_{\tau}\, f$, whose integrability conditions are
$\partial_{\tau}\, ({\hat {\bar n}}_{(a)}\, {}^3{\bar e}_{(a)r}) =
\partial_r\, \sqrt{\sum_a\, {\bar n}^2_{(a)}}$. If these gauges
exist, then the two space-like directions orthogonal to ${\cal {\bar
N}}(\tau ,\vec \sigma )$ (tangent to the 2-surfaces) must be
connected with the two physical degrees of freedom of the
gravitational field in the sense of Ref. \cite{27}. This topic is now under investigation \cite{28}.}.

\medskip

As a consequence, in the canonical formalism we can identify the
following null tetrads (${}^3{\bar e}_{(a)}^r = {\tilde
\phi}^{-1/3}\, Q_a^{-1}\, V_{ra}$)\medskip

\begin{eqnarray*}
 L^A &=&  {1\over {\sqrt{2}}}\, [ l^A - {\cal {\bar N}}^A] = {1\over
{\sqrt{2}}}\, \Big( {1\over {1 + n}};\,\,\,\, - (1 +
{{\sqrt{\sum_c\, {\bar n}^2_{(c)}}}\over {1 + n}})\,\, {\hat {\bar
n}}_{(a)}\, {}^3{\bar e}^r_{(a)} \Big), \nonumber \\
 &&{}\nonumber \\
 K^A &=& {1\over {\sqrt{2}}}\, [ l^A + {\cal {\bar N}}^A] = {1\over
{\sqrt{2}}}\, \Big( {1\over {1 + n}};\,\,\,\, (1 - {{\sqrt{\sum_c\,
{\bar n}^2_{(c)}}}\over {1 + n}})\,\, {\hat {\bar n}}_{(a)}\,
 {}^3{\bar e}^r_{(a)}  \Big), \nonumber \\
 &&{}\nonumber \\
 && \qquad K^{\tau} = L^{\tau},
 \end{eqnarray*}

\bea
 M^A &=& {\cal E}^A_{(+)} =  \Big( 0;\,\,\, {\hat {\bar \epsilon}}_{(+)(a)}\,
{}^3{\bar e}^r_{(a)} \Big) = {\hat {\bar \epsilon}}_{(+)(a)}\,
{}^4{\buildrel \circ \over {\bar E}}^A_{(a)},\nonumber \\
 &&{}\nonumber \\
 M^{*\, A} &=& {\cal E}^A_{(-)} =  \Big( 0;\,\,\,  {\hat
 {\bar \epsilon}}_{(-)(a)}\, {}^3{\bar e}^r_{(a)} \Big) =
 {\hat {\bar \epsilon}}_{(-)(a)}\, {}^4{\buildrel \circ \over
 {\bar E}}^A_{(a)}, \nonumber \\
 &&{}\nonumber \\
 &&{}\nonumber \\
 \sgn\, {}^4g^{AB} &=& L^A\, K^B + L^B\, K^A - (M^A\, M^{*B} + M^B\,
M^{*A}),\nonumber \\
 &&{}\nonumber \\
 0 &= &{}^4g^{AB}\, L_A\, L_B = {}^4g^{AB}\, K_A\, K_B = {}^4g^{AB}\, L_A\, M_B =
{}^4g^{AB}\, K_A\, M_B = {}^4g^{AB}\, M_A\, M_B,\nonumber \\
 &&{}\nonumber \\
 &&{}^4g^{AB}\, L_A\, K_B = \sgn,\qquad {}^4g^{AB}\, M_A\, M^*_B = - \sgn,
 \label{4.4}
 \eea

\noindent whose covariant form  is (${}^3{\bar e}_{(a)}^r = {\tilde
\phi}^{1/3}\, Q_a\, V_{ra}$)\medskip

\bea
 L_A &=& {1\over {\sqrt{2}}}\, \Big(l_A - {\cal {\bar N}}_A\Big)
 = {{\sgn}\over {\sqrt{2}}}\, \Big(1 + n + \sqrt{\sum_c\, {\bar n}^2_{(c)}};
 \,\,\,\, {\hat {\bar n}}_{(a)}\, {}^3{\bar e}_{(a)r}\Big), \nonumber \\
 &&{}\nonumber \\
 K_A &=& {1\over {\sqrt{2}}}\, \Big(l_A + {\cal {\bar N}}_A\Big)
 = {{\sgn}\over {\sqrt{2}}}\, \Big(1 + n - \sqrt{\sum_c\, {\bar n}^2_{(c)}};
 \,\,\,\, - {\hat {\bar n}}_{(a)}\, {}^3{\bar e}_{(a)r}\Big), \nonumber \\
 &&{}\nonumber \\
 M_A &=& - \sgn\, \Big( 0;\,\,\,\,\, {\hat
 {\bar  \epsilon}}_{(+)(a)}\, {}^3{\bar e}_{(a)r}\Big), \nonumber \\
 &&{}\nonumber \\
 M^*_A &=& - \sgn\, \Big( 0;\,\,\,\,\, {\hat
 {\bar  \epsilon}}_{(-)(a)}\, {}^3{\bar e}_{(a)r}\Big).
 \label{4.5}
 \eea

\bigskip

This is a canonical realization of a set of Newman-Penrose null
tetrads \cite{17}.\medskip

Since Eqs.(\ref{2.5}) and (\ref{4.3}) imply

\bea
 {}^4E^A_{(\alpha )} &=& \Big[L^{(o)}{}_{(\alpha )}(\varphi_{(e)})\Big]\, l^A
 + \Big[{\hat {\bar n}}_{(a)}\, R^T_{(a)(b)}(\alpha_{(e)})\,
 L^{(b)}{}_{(\alpha )}(\varphi_{(e)})\Big]\, {\cal {\bar N}} +\nonumber \\
 &+& \Big[{\hat {\bar \epsilon}}_{(-)(a)}\, R^T_{(a)(b)}(\alpha_{(e)})\,
 L^{(b)}{}_{(\alpha )}(\varphi_{(e)})\Big]\, M^A + \Big[{\hat {\bar
 \epsilon}}_{(+)(a)}\, R^T_{(a)(b)}(\alpha_{(e)})\,
 L^{(b)}{}_{(\alpha )}(\varphi_{(e)})\Big]\, M^{*A},\nonumber \\
 &&{}
 \label{4.6}
 \eea

\noindent we see that any other set of null tetrads (built from
arbitrary tetrads ${}^4E^A_{(\alpha )}$) is connected with the null
basis (\ref{4.4}) by Lorentz transformations in the tangent planes
with gauge parameters $\alpha_{(a)}$ and $\varphi_{(a)}$. The
Newman-Penrose Lorentz gauge freedom of the boost, spin and two null
rotations \cite{17} may be re-expressed in terms of these
parameters.
\bigskip

\subsection{The Ricci Scalars and Einstein's Equations}

With the null tetrads (\ref{4.5}) we can replace the energy-momentum tensor ${\hat T}_{AB} = T_{AB} - {1\over 2}\, {}^4g_{AB}\, T$ of Eq.(\ref{2.16}) with the following energy-momentum scalars

 \begin{eqnarray*}
 {\cal T}_{00} &=& {1\over 2}\, {\hat T}_{AB}\, K^A\, K^B =
 {1\over 2}\, T_{AB}\, K^A\, K^B =\nonumber \\
 &=& {1\over 2}\, \Big(T_{(o)(o)} + 2\, T_{(o)(a)}\, {\hat n}_{(a)}
 +T_{(a)(b)}\, {\hat n}_{(a)}\, {\hat n}_{(b)}\Big),\nonumber \\
 {\cal T}_{01} &=& {1\over 2}\, {\hat T}_{AB}\, K^A\, M^B =
 {1\over 2}\, T_{AB}\, K^A\, M^B =\nonumber \\
 &=&    {1\over {\sqrt{2}}}\, {\hat \epsilon}_{(+)(a)}\, \Big(
  T_{(o)(a)} + T_{(a)(b)}\, {\hat n}_{(b)}\Big),\nonumber \\
 {\cal T}_{02} &=& {1\over 2}\, {\hat T}_{AB}\, M^A\, M^B =
 {1\over 2}\, T_{AB}\, M^A\, M^B =\nonumber \\
 &=&    {1\over 2}\, {\hat \epsilon}_{(+)(a)}\,
  {\hat \epsilon}_{(-)(b)}\, T_{(a)(b)},\nonumber \\
 {\cal T}_{11} &=& {1\over 4}\, {\hat T}_{AB}\, (K^A\, L^B + M^a\, M^{*\, B}) =
 {1\over 4}\, T_{AB}\, (K^A\, L^B + M^A\, M^{*\, B}) =\nonumber \\
 &=&   {1\over 2}\, \Big(T_{(o)(o)} + (2\, {\hat \epsilon}_{(+)(a)}\,
 {\hat \epsilon}_{(-)(b)} - {\hat n}_{(a)}\, {\hat n}_{(b)})\,
 T_{(a)(b)}\Big),\nonumber \\
 {\cal T}_{12} &=& {1\over 2}\, {\hat T}_{AB}\, L^A\, M^B =
 {1\over 2}\, T_{AB}\, L^A\, M^B =\nonumber \\
 &=&    {1\over {\sqrt{2}}}\, {\hat \epsilon}_{(+)(a)}\, \Big(
  T_{(o)(a)} - T_{(a)(b)}\, {\hat n}_{(b)}\Big),\nonumber \\
 {\cal T}_{22} &=& {1\over 2}\, {\hat T}_{AB}\, L^A\, L^B =
 {1\over 2}\, T_{AB}\, L^A\, L^B =\nonumber \\
 &=&    {1\over 2}\, \Big(T_{(o)(o)} - 2\, T_{(o)(a)}\, {\hat n}_{(a)}
 +T_{(a)(b)}\, {\hat n}_{(a)}\, {\hat n}_{(b)}\Big),\nonumber \\
 {\cal T} &=& {{\sgn}\over 4}\, {}^4g^{AB}\, {\hat T}_{AB} = - {{\sgn}\over 4}\, T,
 \end{eqnarray*}

\bea
 T_{AB} &=& 2\, \Big[{\cal T}_{oo}\, L_A\, L_B + {\cal T}_{22}\, K_A\, K_B + 2\, {\cal T}_{11}\,
 (L_A\, K_B + L_B\, K_A) -\nonumber \\
 &-& {\cal T}_{o1}\, (L_A\, M_B^* + L_B\, M_A^*) - {\cal T}^*_{o1}\,
 (L_A\, M_B + L_B\, M_A) -\nonumber \\
 &-& {\cal T}_{12}\, (K_A\, M_B^* + K_B\, M_A^*) - {\cal T}^*_{12}\, (K_A\, M_B + K_B\, M_A)
 +\nonumber \\
 &+& {\cal T}_{o2}\, M_A^*\, M_B^* + {\cal T}_{o2}^*\, M_A\, M_B  \Big] + ({\cal T}_{11} +
 {\cal T})\, (M_A\, M_B^* + M_B\, M_A^*),\nonumber \\
 {}&&\nonumber \\
 {\hat T}_{AB} &=& T_{AB} - {1\over 2}\, {}^4g_{AB}\, T = T_{AB} + 2\,
 [L_A\, K_B + L_B\, K_A - (M_A\, M_B^* + M_A^*\, M_B)]\, {\cal T}.\nonumber \\
 {}&&
  \label{4.7}
 \eea

\bigskip

The null tetrads (\ref{4.5}) allow us to get the following
expressions for the 9 Ricci scalars ($\Phi_{01}$, $\Phi_{02}$ and
$\Phi_{12}$ are complex numbers) and the curvature scalar (${}^4R =
24\, \Lambda$) replacing the 10 components of the 4-Ricci tensor in the Newman-Penrose formalism \cite{17}
\bigskip

\bea
 \Phi_{00} &=& {1\over 2}\, {}^4R_{AB}\, K^A\, K^B\,\, = {{4\pi\,
 G}\over {c^3}}\, {\cal T}_{00} + {1\over 2}\, {\cal E}_{00}
 \cir\,\, {{4\pi\, G}\over {c^3}}\, {\cal T}_{oo},\nonumber \\
  \Phi_{01} &=& {1\over 2}\, {}^4R_{AB}\, K^A\, M^B\,\, = {{4\pi\,
 G}\over {c^3}}\, {\cal T}_{01} + {1\over 2}\, {\cal E}_{01}
  \cir\,\, {{4\pi\, G}\over {c^3}}\, {\cal T}_{01}, \nonumber \\
  \Phi_{02} &=& {1\over 2}\, {}^4R_{AB}\, M^A\, M^B\,\, = {{4\pi\,
 G}\over {c^3}}\, {\cal T}_{02} + {1\over 2}\, {\cal E}_{02}
  \cir\,\, {{4\pi\, G}\over {c^3}}\, {\cal T}_{02},\nonumber \\
 \Phi_{11} &=& {1\over 4}\, {}^4R_{AB}\, (K^A\,
 L^B + M^A\, M^{*B})\,\, = {{2\pi\,
 G}\over {c^3}}\, {\cal T}_{11} + {1\over 4}\, {\cal E}_{11}
 \cir {{2\pi\, G}\over {c^3}}\, {\cal T}_{11},\nonumber \\
  \Phi_{12} &=& {1\over 2}\, {}^4R_{AB}\, L^A\, M^B\,\, = {{4\pi\,
 G}\over {c^3}}\, {\cal T}_{12} + {1\over 2}\, {\cal E}_{12}
  \cir\,\, {{4\pi\, G}\over {c^3}}\, {\cal T}_{12},\nonumber \\
  \Phi_{22} &=& {1\over 2}\, {}^4R_{AB}\, L^A\, L^B\,\, = {{4\pi\,
 G}\over {c^3}}\, {\cal T}_{22} + {1\over 2}\, {\cal E}_{22}
   \cir\,\, {{4\pi\, G}\over {c^3}}\, {\cal T}_{22},\nonumber \\
 6\, \sgn\, \Lambda &=&
 {{\sgn}\over 4}\, {}^4R = {1\over 2}\, {}^4R_{AB}\, (K^A\, L^B - M^A\, M^{*B})\,\,
 = - \sgn\, \Big({{2\pi\, G}\over {c^3}}\, {\cal T} + {1\over 4}\, {\cal E}\Big)
 \cir\,\, - \sgn\, {{2\pi\, G}\over {c^3}}\, {\cal T},\nonumber \\
 &&{}
 \label{4.8}
\eea

\bigskip

In Eqs.(\ref{4.8}) we introduced the following notation for Einstein's equations (\ref{2.16})

\begin{eqnarray*}
 &&{\cal E}_{00} = {1\over 2}\, E_{AB}\, K^A\, K^B = \Phi_{00} -
 {{8\pi G}\over {c^3}}\, {\cal T}_{00} \cir 0,\nonumber \\
 &&{\cal E}_{01} = {1\over 2}\, E_{AB}\, K^A\, M^B = \Phi_{01} -
 {{8\pi G}\over {c^3}}\, {\cal T}_{01} \cir 0,\nonumber \\
 &&{\cal E}_{02} = {1\over 2}\, E_{AB}\, M^A\, M^B = \Phi_{02} -
 {{8\pi G}\over {c^3}}\, {\cal T}_{02} \cir 0,\nonumber \\
 &&{\cal E}_{11} = {1\over 4}\, E_{AB}\, (K^A\, L^B + M^A\, M^{*\, B}) = \Phi_{11} -
 {{8\pi G}\over {c^3}}\, {\cal T}_{11} \cir 0,\nonumber \\
 &&{\cal E}_{12} = {1\over 2}\, E_{AB}\, L^A\, M^B = \Phi_{12} -
 {{8\pi G}\over {c^3}}\, {\cal T}_{12} \cir 0,\nonumber \\
 &&{\cal E}_{22} = {1\over 2}\, E_{AB}\, L^A\, L^B = \Phi_{22} -
 {{8\pi G}\over {c^3}}\, {\cal T}_{22} \cir 0,\nonumber \\
 &&{\cal E} = {{\sgn}\over 4}\, {}^4g^{AB}\, E_{AB} = 6\, \epsilon\,
 \Lambda - {{8\pi G}\over {c^3}}\, {\cal T} \cir 0,
 \end{eqnarray*}

\bea
 E_{AB} &=& 2\, \Big[{\cal E}_{oo}\, L_A\, L_B + {\cal E}_{22}\, K_A\, K_B + 2\, {\cal E}_{11}\,
 (L_A\, K_B + L_B\, K_A) -\nonumber \\
 &-& {\cal E}_{o1}\, (L_A\, M_B^* + L_B\, M_A^*) - {\cal E}^*_{o1}\,
 (L_A\, M_B + L_B\, M_A) -\nonumber \\
 &-& {\cal E}_{12}\, (K_A\, M_B^* + K_B\, M_A^*) - {\cal E}^*_{12}\, (K_A\, M_B + K_B\, M_A)
 +\nonumber \\
 &+& {\cal E}_{o2}\, M_A^*\, M_B^* + {\cal E}_{o2}^*\, M_A\, M_B  \Big] + ({\cal E}_{11} +
 {\cal E})\, (M_A\, M_B^* + M_B\, M_A^*).\nonumber \\
 {}&&
 \label{4.9}
 \eea

The combinations (\ref{2.19}) of Einstein equations corresponding to the
Hamiltonian super-Hamiltonian and super-momentum constraints now can be written in the following form
(${}^3{\bar e}^r_{(a)} = {\tilde \phi}^{-1/3}\, Q_a^{-1}\, V_{ra}$)

\begin{eqnarray*}
 &&\Phi_{00} + \Phi_{22} + 2\, \Phi_{11} - 6\, \sgn\, \Lambda
 \approx\nonumber \\
 &&{}\nonumber \\
 &&\qquad \approx {{8\pi\, G}\over {c^3\, (1 + n)^2}}\,
 \Big[T_{\tau\tau} - 2\, T_{\tau r}\, {}^3{\bar e}^r_{(a)}\, {\bar n}_{(a)} +
 T_{rs}\, {}^3{\bar e}^r_{(a)}\, {}^3{\bar e}^s_{(b)}\, {\bar n}_{(a)}\,
 {\bar n}_{(b)}\Big],
 \end{eqnarray*}

 \bea
 &&\Phi_{00} - \Phi_{22} \approx \sgn\, {{8\pi\, G}\over {c^3\,
 (1 + n)}}\, {\hat {\bar n}}_{(a)}\, {}^3{\bar e}^r_{(a)}\,
 \Big[T_{\tau r} - T_{rs}\, {}^3{\bar e}^s_{(b)}\, {\bar n}_{(b)}\Big],\nonumber \\
 &&\Phi_{01} - \Phi_{12} \approx \sgn\, {{4\sqrt{2}\, \pi\, G}\over {c^3\,
 (1 + n)}}\, {\hat {\bar \epsilon}}_{(+)(a)}\, {}^3{\bar e}^r_{(a)}\,
 \Big[T_{\tau r} - T_{rs}\, {}^3{\bar e}^s_{(b)}\, {\bar n}_{(b)}\Big],\nonumber \\
   &&\Phi_{01}^* - \Phi_{12}^* \approx \sgn\, {{4\sqrt{2}\, \pi\, G}\over {c^3\,
 (1 + n)}}\, {\hat {\bar \epsilon}}_{(-)(a)}\, {}^3{\bar e}^r_{(a)}\,
 \Big[T_{\tau r} - T_{rs}\, {}^3{\bar e}^s_{(b)}\, {\bar
 n}_{(b)}\Big].
 \label{4.10}
 \eea

We have the following inversion formulas

\begin{eqnarray*}
 {}^4R_{AB} &=& 2 \Phi_{00}\, L_A\, L_B + 2\, \Phi_{22}\, K_A\, K_B +
 2\, \Phi_{02}\, M_A^*\, M_B^* + 2\, \Phi^*_{02}\, M_A\, M_B -\nonumber \\
 &-& 2\, \Phi_{01}\, (L_A\, M_B^* + L_B\, M_A^*) - 2\, \Phi^*_{01}\,
 (L_A\, M_B + L_B\, M_A) -\nonumber \\
 &-& 2\, \Phi_{12}\, (K_A\, M_B^* + K_B\,
 M_A^*) - 2\, \Phi^*_{12}\, (K_A\, M_B + K_B\, M_A) +\nonumber \\
 &+& 2\, \Phi_{11}\, (L_A\, K_B + L_B\, K_A + M_A\, M_B^* + M_B\, M_A^*)
 +\nonumber \\
 &+& 6\,\sgn\, \Lambda\, [L_A\, K_B + L_B\, K_A -(M_A\, M_B^* + M_B\,
 M_A^*)],
 \end{eqnarray*}

\begin{eqnarray*}
 {}^4R_{\tau\tau} &=& (1 + n + \sqrt{\sum_c\, {\bar n}^2_{(c)}}\,\,)^2\, \Phi_{00}
 + (1 + n - \sqrt{\sum_c\, {\bar n}^2_{(c)}}\,\,)^2\, \Phi_{22} +\nonumber \\
  &+&  2\, [(1 + n)^2 - {\bar n}_{(a)}\, {\bar n}_{(a)}]\, (\Phi_{11} +
   3\, \sgn\,  \Lambda),
  \end{eqnarray*}

\begin{eqnarray*}
  {}^4R_{\tau r} &=& {\tilde \phi}^{-1/3}\, Q_a^{-1}\, V_{ra}
  \, \Big[{\hat {\bar n}}_{(a)}\, \Big((1 + n +
  \sqrt{\sum_c\, {\bar n}^2_{(c)}}\,\,)\, \Phi_{00} - (1 + n -
  \sqrt{\sum_c\, {\bar n}^2_{(c)}}\,\,)\, \Phi_{22} -\nonumber \\
 &-& 2\, \sqrt{\sum_c\, {\bar n}^2_{(c)}}\, (\Phi_{11} +
 3\, \sgn\, \Lambda)\Big) +\nonumber \\
 &+& \sqrt{2}\, \Big((1 + n + \sqrt{\sum_c\, {\bar n}^2_{(c)}}\,\,)\, \Phi^*_{01}
 - (1 + n - \sqrt{\sum_c\, {\bar n}^2_{(c)}}\,\,)\, \Phi_{12}^*\Big)\,
 {\hat {\bar \epsilon}}_{(+)(a)} +\nonumber \\
 &+& \sqrt{2}\, \Big((1 + n + \sqrt{\sum_c\, {\bar n}^2_{(c)}}\,\,)\, \Phi_{01}
 - (1 + n - \sqrt{\sum_c\, {\bar n}^2_{(c)}}\,\,)\, \Phi_{12}\Big)\,
 {\hat {\bar \epsilon}}_{(-)(a)}\Big],
\end{eqnarray*}

\bea
  {}^4R_{rs} &=& {\tilde \phi}^{2/3}\, Q_a\, Q_b\, V_{ra}\, V_{sb}\, \Big[{\hat
  {\bar n}}_{(a)}\,  {\hat {\bar n}}_{(b)}\, (\Phi_{00} + \Phi_{22}) +\nonumber \\
  &+& 2\, {\hat {\bar \epsilon}}_{(-)(a)}\, {\hat {\bar \epsilon}}_{(-)(b)}\,
  \Phi_{02} + 2\, {\hat {\bar \epsilon}}_{(+)(a)}\, {\hat {\bar \epsilon}}_{(+)(b)}\,
  \Phi^*_{02} +\nonumber \\
  &+& \sqrt{2}\, ({\hat {\bar n}}_{(a)}\, {\hat {\bar \epsilon}}_{(-)(b)} +
  {\hat {\bar n}}_{(b)}\, {\hat {\bar \epsilon}}_{(-)(a)})\, (\Phi_{01} -
  \Phi_{12}) +\nonumber \\
  &+& \sqrt{2}\, ({\hat {\bar n}}_{(a)}\, {\hat {\bar \epsilon}}_{(+)(b)} +
  {\hat {\bar n}}_{(b)}\, {\hat {\bar \epsilon}}_{(+)(a)})\, (\Phi^*_{01} -
  \Phi^*_{12}) -\nonumber \\
  &-& 2\, ({\hat {\bar n}}_{(a)}\, {\hat {\bar n}}_{(b)} - {\hat {\bar  \epsilon}}_{(+)(a)}\,
  {\hat {\bar \epsilon}}_{(-)(b)} - {\hat {\bar \epsilon}}_{(+)(b)}\,
  {\hat {\bar \epsilon}}_{(-)(a)})\, \Phi_{11} -\nonumber \\
  &-& 6\, \sgn\, \delta_{(a)(b)}\, \Lambda  \Big].
 \label{4.11}
 \eea

 \bigskip

The first of Eqs.(\ref{3.6}) can now be written in the following form

\bea
 {}^4R_{\tau r\tau s} &=& {\bar W}_{\tau r\tau s} + \sgn\, (1 + n)^2\,
 {\tilde \phi}^{2/3}\, Q_a\, Q_b\, V_{ra}\, V_{sb}\, \Big[{\hat
  {\bar n}}_{(a)}\,  {\hat {\bar n}}_{(b)}\, (\Phi_{00} + \Phi_{22}) +\nonumber \\
  &+& 2\, {\hat {\bar \epsilon}}_{(-)(a)}\, {\hat {\bar \epsilon}}_{(-)(b)}\,
  \Phi_{02} + 2\, {\hat {\bar \epsilon}}_{(+)(a)}\, {\hat {\bar \epsilon}}_{(+)(b)}\,
  \Phi^*_{02} +\nonumber \\
  &+& \sqrt{2}\, ({\hat {\bar n}}_{(a)}\, {\hat {\bar \epsilon}}_{(-)(b)} +
  {\hat {\bar n}}_{(b)}\, {\hat {\bar \epsilon}}_{(-)(a)})\, (\Phi_{01} -
  \Phi_{12}) +\nonumber \\
  &+& \sqrt{2}\, ({\hat {\bar n}}_{(a)}\, {\hat {\bar \epsilon}}_{(+)(b)} +
  {\hat {\bar n}}_{(b)}\, {\hat {\bar \epsilon}}_{(+)(a)})\, (\Phi^*_{01} -
  \Phi^*_{12}) -\nonumber \\
  &-& 2\, ({\hat {\bar n}}_{(a)}\, {\hat {\bar n}}_{(b)} - {\hat {\bar  \epsilon}}_{(+)(a)}\,
  {\hat {\bar \epsilon}}_{(-)(b)} - {\hat {\bar \epsilon}}_{(+)(b)}\,
  {\hat {\bar \epsilon}}_{(-)(a)})\, \Phi_{11} -\nonumber \\
  &-& 6\, \sgn\, \delta_{(a)(b)}\, \Lambda  \Big],
 \label{4.12}
  \eea

\noindent and it can be written in terms of the null tetrad expressions of the energy-momentum tensor
and of Einstein's equations by using Eq.(\ref{4.8}).

\section{Conclusions}

We have found that in the framework of canonical ADM tetrad gravity in the York canonical basis there is a Hamiltonian radar tensor, which coincides with the 4-Riemann tensor on the solutions of Einstein's equations. Therefore "on-shell" there is a Hamiltonian 4-Riemann radar tensor, whose components are 4-scalars of the space-time due to the use of radar 4-coordinates.

\medskip

Moreover, the 3+1 splitting of the space-time used to define the phase space allows us to introduce Hamiltonian null tetrads. This opens the way to get a Hamiltonian formulation of the Newman-Penrose formalism. We have given the 4-Ricci scalars as the sum of Einstein's equations and of terms depending on the Hamiltonian expression of the energy-momentum tensor of the matter.

\medskip

In the second paper these results will be used to get the Hamiltonian expression of the 4-Weyl tensor, of the 4-Weyl scalars and of the four 4-Weyl eigenvalues. Moreover we will discuss the problem of the determination of the DO's and BO's of the gravitational field. The Weyl eigenvalues will be shown to be neither DO's nor BO's, but only the 4-scalars
needed to give a physical identification as point-events of the mathematical points of the space-time 4-manifold \cite{7}.

\appendix

\section{The 3-Riemann and 3-Ricci Tensors in the York Canonical Basis}

The 3-Riemann  and 3-Ricci tensors sare

\begin{eqnarray*}
 {}^3R^r{}_{suv} &=& \partial_u\, {}^3\Gamma^r_{sv} -
\partial_v\, {}^3\Gamma^r_{su} + {}^3\Gamma^h_{sv}\,
{}^3\Gamma^r_{hu} - {}^3\Gamma^h_{su}\,
{}^3\Gamma^r_{hv},\nonumber \\
 {}^3R_{rsuv} &=& {1\over 2}\, \Big(\partial_r\, \partial_u\,
{}^3g_{sv} + \partial_s\, \partial_v\, {}^3g_{ru} - \partial_r\,
\partial_v\, {}^3g_{su} - \partial_s\, \partial_u\, {}^3g_{rv}\Big)
+\nonumber \\
 &+& {}^3g_{hk}\, \Big({}^3\Gamma^h_{rv}\,
{}^3\Gamma^k_{su} - {}^3\Gamma^h_{ru}\, {}^3\Gamma^k_{sv}\Big),
\nonumber \\
 &&{}\nonumber \\
 {}^3R_{rs} &=& {}^3g^{uv}\, {}^3R_{urvs},\qquad {}^3R =
 {}^3g^{rs}\, {}^3R_{rs},
\end{eqnarray*}

\begin{eqnarray*}
 {}^3R_{rsuv} &=& {1\over 2}\, {\tilde \phi}^{2/3}\, Q^2_a\nonumber \\
 &&\Big\{\,\, {2\over 3}\, \Big[\Big(\partial_r\, ({\tilde \phi}^{-1}\,
 \partial_u\, \tilde \phi)\, V_{sa} - \partial_s\, ({\tilde \phi}^{-1}\,
 \partial_u\, \tilde \phi)\, V_{ra}\Big)\, V_{va} +\nonumber \\
 &+& \Big(\partial_s\, ({\tilde \phi}^{-1}\, \partial_v\, \tilde \phi)\, V_{ra}
 - \partial_r\, ({\tilde \phi}^{-1}\, \partial_v\, \tilde \phi)\, V_{sa}\Big)\,
 V_{ua}\Big] +\nonumber \\
 &+& 2\, \Big((\partial_r\, \partial_u\, \Gamma_a^{(1)}\, V_{sa} -
 \partial_s\, \partial_u\, \Gamma_a^{(1)}\, V_{ra})\, V_{va} +
 (\partial_s\, \partial_v\, \Gamma_a^{(1)}\, V_{ra} -
 \partial_r\, \partial_v\, \Gamma_a^{(1)}\, V_{sa})\, V_{ua}
 \Big) +\nonumber \\
 &+& 4\, \Big[({1\over 3}\, {\tilde \phi}^{-1}\, \partial_r\, \tilde \phi
 + \partial_r\, \Gamma_a^{(1)})\, V_{sa} - ({1\over 3}\, {\tilde \phi}^{-1}\,
 \partial_s\, \tilde \phi + \partial_s\, \Gamma_a^{(1)})\, V_{ra}\Big]\nonumber \\
 &&\Big[({1\over 3}\, {\tilde \phi}^{-1}\, \partial_u\, \tilde \phi
 + \partial_u\, \Gamma_a^{(1)})\, V_{va} - ({1\over 3}\, {\tilde \phi}^{-1}\,
 \partial_v\, \tilde \phi + \partial_v\, \Gamma_a^{(1)})\, V_{ua}\Big] +\nonumber \\
 &+& 2\, \Big[({1\over 3}\, {\tilde \phi}^{-1}\, \partial_r\, \tilde \phi
 + \partial_r\, \Gamma_a^{(1)})\, \Big(\partial_u\, (V_{sa}\, V_{va}) -
 \partial_v\, (V_{sa}\, V_{ua})\Big) +\nonumber \\
 &+& ({1\over 3}\, {\tilde \phi}^{-1}\, \partial_s\, \tilde \phi
 + \partial_s\, \Gamma_a^{(1)})\, \Big(\partial_v\, (V_{ra}\, V_{ua}) -
 \partial_u\, (V_{ra}\, V_{va})\Big) +\nonumber \\
 &+& ({1\over 3}\, {\tilde \phi}^{-1}\, \partial_u\, \tilde \phi
 + \partial_u\, \Gamma_a^{(1)})\, \Big(\partial_r\, (V_{sa}\, V_{va}) -
 \partial_s\, (V_{ra}\, V_{va})\Big) +\nonumber \\
 &+& ({1\over 3}\, {\tilde \phi}^{-1}\, \partial_v\, \tilde \phi
 + \partial_v\, \Gamma_a^{(1)})\, \Big(\partial_s\, (V_{ra}\, V_{ua}) -
 \partial_r\, (V_{sa}\, V_{ua})\Big) \Big] +\nonumber \\
 &+& \partial_r\, \partial_u\, (V_{sa}\, V_{va}) - \partial_r\,
 \partial_v\, (V_{sa}\, V_{ua}) + \partial_s\, \partial_v\,
 (V_{ra}\, V_{ua}) - \partial_s\, \partial_u\, (V_{ra}\, V_{va}) -
 \nonumber \\
 &-& {1\over 9}\, \Big[(V_{ra}\, {\tilde \phi}^{-1}\, \partial_v\, \tilde \phi
 + V_{va}\, {\tilde \phi}^{-1}\, \partial_r\, \tilde \phi)\,
 (V_{sa}\, {\tilde \phi}^{-1}\, \partial_u\, \tilde \phi +
 V_{ua}\, {\tilde \phi}^{-1}\, \partial_s\, \tilde \phi) +\nonumber \\
 &+&Q_b^2\, Q_d^2\, Q_a^{-4}\, V_{rb}\, V_{vb}\, V_{sd}\, V_{ud}\, V_{ta}\, V_{wa}\,
 {\tilde \phi}^{-1}\, \partial_t\, \tilde \phi\,
 {\tilde \phi}^{-1}\, \partial_w\, \tilde \phi   -\nonumber \\
 &-& Q_b^2\, Q_a^{-2}\, V_{ta}\, {\tilde \phi}^{-1}\, \partial_t\, \tilde \phi\,
 \Big(V_{sb}\, V_{ub}\, (V_{ra}\, {\tilde \phi}^{-1}\, \partial_v\, \tilde \phi
 + V_{va}\, {\tilde \phi}^{-1}\, \partial_r\, \tilde \phi) +\nonumber \\
 &+& V_{rb}\, V_{vb}\, (V_{sa}\, {\tilde \phi}^{-1}\, \partial_u\, \tilde \phi
 + V_{ua}\, {\tilde \phi}^{-1}\, \partial_s\, \tilde \phi)\Big)\Big] -\nonumber \\
  &-& {1\over 3}\, \Big[(V_{ra}\, {\tilde \phi}^{-1}\, \partial_v\, \tilde \phi +
 V_{va}\, {\tilde \phi}^{-1}\, \partial_r\, \tilde \phi)\nonumber \\
 &&\sum_{\bar a}\, [\gamma_{\bar aa}\, (V_{sa}\, \partial_u\, R_{\bar a} +
 V_{ua}\, \partial_s\, R_{\bar a}) - \gamma_{\bar ab}\, Q_b^2\, Q_a^{-2}\,
 V_{sb}\, V_{ub}\, V_{ta}\, \partial_t\, R_{\bar a}] +\nonumber \\
 \end{eqnarray*}

\begin{eqnarray*}
 &+& (V_{sa}\, {\tilde \phi}^{-1}\, \partial_u\, \tilde \phi +
 V_{ua}\, {\tilde \phi}^{-1}\, \partial_s\, \tilde \phi)\nonumber \\
 &&\sum_{\bar a}\, [\gamma_{\bar aa}\, (V_{ra}\, \partial_v\, R_{\bar a} +
 V_{va}\, \partial_r\, R_{\bar a}) - \gamma_{\bar ab}\, Q_b^2\, Q_a^{-2}\,
 V_{rb}\, V_{vb}\, V_{ta}\, \partial_t\, R_{\bar a}] -\nonumber \\
 &-&Q_b^2\, Q_a^{-2}\, V_{ta}\, {\tilde \phi}^{-1}\, \partial_t\, \tilde
 \phi\, \sum_{\bar a}\, \gamma_{\bar aa}\nonumber \\
 &&[V_{rb}\, V_{vb}\, (V_{sa}\, \partial_u\, R_{\bar a} + V_{ua}\,
 \partial_s\, R_{\bar a}) + V_{sb}\, V_{ub}\, (V_{ra}\, \partial_v\,
 R_{\bar a} + V_{va}\, \partial_r\, R_{\bar a})] -\nonumber \\
 &-& Q_b^2\, Q_d^2\, Q_a^{-4}\, \sum_{\bar a}\, \gamma_{\bar ab}\,
 [V_{sb}\, V_{ub}\, V_{rd}\, V_{vd} + V_{rb}\, V_{vb}\, V_{sd}\, V_{ud}]\,
 V_{ta}\, V_{wa}\, {\tilde \phi}^{-1}\, \partial_t\, \tilde
 \phi\, \partial_w\, R_{\bar a}\Big] -\nonumber \\
 &-& \sum_{\bar a\bar b}\, \Big[\gamma_{\bar aa}\, (V_{ra}\, \partial_v\, R_{\bar a}
 + V_{va}\, \partial_r\, R_{\bar a}) - \gamma_{\bar ab}\, Q_b^2\, Q_a^{-2}\, V_{rb}\,
 V_{vb}\, V_{ta}\, \partial_t\, R_{\bar a}\Big]\nonumber \\
 &&\Big[\gamma_{\bar ba}\, (V_{sa}\, \partial_u\, R_{\bar b} + V_{ua}\, \partial_s\, R_{\bar b})
 - \gamma_{\bar bd}\, Q_d^2\, Q_a^{-2}\, V_{sd}\, V_{ud}\, V_{wa}\, \partial_w\, R_{\bar b}\Big] -
 \nonumber \\
 &-& {1\over 6}\, \Big[[V_{ra}\, {\tilde \phi}^{-1}\, \partial_v\, \tilde \phi +
 V_{va}\, {\tilde \phi}^{-1}\, \partial_r\, \tilde \phi - Q_b^2\, Q_a^{-2}\,
 V_{rb}\, V_{vb}\, V_{ta}\, {\tilde \phi}^{-1}\, \partial_t\, \tilde \phi]\nonumber \\
 &&\Big[\partial_s\, V_{ua} + \partial_u\, V_{sa} + Q_d^2\, Q_a^{-2}\, V_{wa}\,
 [V_{sd}\, (\partial_u\, V_{wd} - \partial_w\, V_{ud}) + V_{ud}\, (\partial_s\,
 V_{wd} - \partial_w\, V_{sd})]\Big] +\nonumber \\
 &+& [V_{sa}\, {\tilde \phi}^{-1}\, \partial_u\, \tilde \phi + V_{ua}\,
 {\tilde \phi}^{-1}\, \partial_s\, \tilde \phi - Q_b^2\, Q_a^{-2}\, V_{sb}\, V_{ub}\, V_{ta}\,
 {\tilde \phi}^{-1}\, \partial_t\, \tilde \phi]\nonumber \\
 &&\Big[\partial_r\, V_{va} + \partial_v\, V_{ra} + Q_d^2\, Q_a^{-2}\, V_{wa}\,
 [V_{rd}\, (\partial_v\, V_{wd} - \partial_w\, V_{vd}) + V_{vd}\, (\partial_r\,
 V_{wd} - \partial_w\, V_{rd})]\Big]\Big] -\nonumber \\
 &-& {1\over 2}\, \sum_{\bar a}\, \Big[[\gamma_{\bar aa}\, (V_{ra}\, \partial_v\,
 R_{\bar a} + V_{va}\, \partial_r\, R_{\bar a}) - \gamma_{\bar ab}\, Q_b^2\, Q_a^{-2}\,
 V_{rb}\, V_{vb}\, V_{ta}\, \partial_t\, R_{\bar a}]\nonumber \\
 &&\Big[\partial_s\, V_{ua} + \partial_u\, V_{sa} + Q_d^2 Q_a^{-2}\, V_{wa}\, [V_{sd}\,
 (\partial_u\, V_{wd} - \partial_w\, V_{ud}) + V_{ud}\, (\partial_s\, V_{wd} -
 \partial_w\, V_{sd})]\Big] +\nonumber \\
 &+& [\gamma_{\bar aa}\, (V_{sa}\, \partial_u\, R_{\bar a} + V_{ua}\, \partial_s\, R_{\bar a})
 - \gamma_{\bar ab}\, Q_b^2\, Q_a^{-2}\, V_{sb}\, V_{ub}\, V_{ta}\, \partial_t\, R_{\bar a}]\nonumber \\
 &&\Big[\partial_r\, V_{va} + \partial_v\, V_{ra} + Q_d^2\, Q_a^{-2}\, V_{wa}\,
 [V_{rd}\, (\partial_v\, V_{wd} - \partial_w\, V_{vd}) + V_{vd}\, (\partial_r\,
 V_{wd} - \partial_w\, V_{rd})]\Big]\Big] -\nonumber \\
 &-& {1\over 4}\, \Big[\partial_r\, V_{va} + \partial_v\, V_{ra} + Q_b^2\, Q_a^{-2}\, V_{ta}\,
 [V_{rb}\, (\partial_v\, V_{tb} - \partial_t\, V_{vb}) + V_{vb}\, (\partial_r\,
 V_{tb} - \partial_t\, V_{rb})]\Big]\nonumber \\
 &&\Big[\partial_s\, V_{ua} + \partial_u\, V_{sa} + Q_d^2\, Q_a^{-2}\, V_{wa}\,
 [V_{sd}\, (\partial_u\, V_{wd} - \partial_w\, V_{ud}) + V_{ud}\, (\partial_s\,
 V_{wd} - \partial_w\, V_{sd})]\Big]\, \Big\},
 \end{eqnarray*}

\begin{eqnarray*}
 {}^3R_{sv} &=& -\sgn \sum_{acd}{Q^2_a \over Q^2_c Q^2_d} \Big( \delta_{(c)(d)} - {\bar{n}_{(c)} \bar{n}_{(d)} \over (1+n)^2}\Big) \, {1 \over 2}\\ \nonumber
&& \Big\{ {2\over 3} \Big[ \Big( V_{rc} \partial_r (\tilde \phi^{-1} \partial_u \tilde \phi) V_{sa} - \partial_s (\tilde \phi^{-1} \partial_u \tilde \phi) \delta_{ac} \Big) V_{va} V_{ud} + \\ \nonumber
&& + \Big( \partial_s (\tilde \phi^{-1} \partial_v \tilde \phi) \delta_{ac} - V_{rc} \partial_r (\tilde \phi^{-1} \partial_v \tilde \phi) V_{sa} \Big) \delta_{ad} \Big] + \\ \nonumber
&& + 2 \Big( (V_{rc} \partial_r \partial_u \Gamma^{(1)}_a V_{sa} - \partial_s \partial_u \Gamma^{(1)}_a \delta_{ac}) V_{va} V_{ud} + (\partial_s \partial_v \Gamma^{(1)}_a \delta_{ac} - V_{rc}\partial_r \partial_v \Gamma^{(1)}_a V_{sa}) \delta_{ad} V_{ud}\Big) + \\ \nonumber
 \end{eqnarray*}

\begin{eqnarray*}
&& + 4 \Big[ \Big( {1 \over 3} \tilde \phi^{-1} \partial_r \tilde \phi + \partial_r \Gamma_a \Big) V_{rc} V_{sa} - \Big( {1 \over 3} \tilde \phi^{-1} \partial_s \tilde \phi + \partial_s \Gamma^{(1)}_a \Big) \delta_{ac} \Big] \\ \nonumber
&& \Big[ \Big( {1 \over 3} \tilde \phi^{-1} \partial_u \tilde \phi + \partial_u \Gamma^{(1)}_a\Big) V_{va}V_{ud} - \Big( {1 \over 3} \tilde \phi^{-1} \partial_v \tilde \phi + \partial_v \Gamma^{(1)}_a \Big) \delta_{ac} \Big]+ \\ \nonumber
&&+ 2 V_{rc} V_{vd} \Big[ \Big( {1 \over 3} \tilde \phi^{-1} \partial_r \tilde \phi + \partial_r \Gamma^{(1)}_a \Big) \Big( \partial_u (V_{sa}V_{va}) - \partial_v(V_{sa} V_{ua}) \Big) + \\ \nonumber
&&+\Big( {1 \over 3} \tilde \phi^{-1} \partial_s \tilde \phi + \partial_s \Gamma^{(1)}_a \Big) \Big( \partial_v(V_{ra}V_{ua} - \partial_u(V_{ra} V_{va}) \Big) + \\ \nonumber
&&+\Big( {1 \over 3} \tilde \phi^{-1} \partial_u \tilde \phi  + \partial_u \Gamma^{(1)}_a\Big) \Big( \partial_r (V_{sa} V_{va}) - \partial_s(V_{ra} V_{va}) \Big) + \\ \nonumber
&&+ \Big( {1 \over 3} \tilde \phi^{-1} \partial_v \tilde \phi + \partial_v \Gamma^{(1)}_a \Big) \Big( \partial_s(V_{ra} V_{ua}) - \partial_r(V_{va} V_{ua}) \Big)\Big] + \\ \nonumber
&&+ V_{rc} V_{ud} \Big[ \partial_r \partial_u(V_{sa}V_{va}) - \partial_r \partial_v(V_{sa}V_{ua}) + \partial_s \partial_v (V_{ra}V_{ua}) - \partial_s \partial_u (V_{ra} V_{va}) \Big] - \\ \nonumber
&-& {1 \over 9} \Big[ \Big( \delta_{ac}(\tilde \phi^{-1} \partial_v \tilde \phi) + V_{rc} V_{va}(\tilde \phi^{-1} \partial_r \tilde \phi) \Big) \Big( V_{sa} V_{ud} (\tilde \phi^{-1} \partial_u \tilde \phi) + \delta_{ad} (\tilde \phi^{-1} \partial_s \tilde \phi) \Big) + \\ \nonumber
&+& {Q^2_c Q^2_d \over Q_a^4} V_{vc} V_{sd} V_{ta} V_{wa} (\tilde \phi^{-1} \partial_t \tilde \phi) (\tilde \phi^{-1} \partial_w \tilde \phi) + \\ \nonumber
&-& {Q^2_b \over Q^2_a} V_{ta} (\tilde \phi^{-1} \partial_t \tilde \phi) \Big( V_{sb} V_{ub} (V_{ra} \tilde \phi^{-1} \partial_v \tilde \phi + V_{va} \tilde \phi^{-1} \partial_r \tilde \phi) + V_{rb} V_{vb} (V_{sa} \tilde \phi^{-1} \partial_u \tilde \phi + V_{ua} \tilde \phi^{-1} \partial_s \tilde \phi)\Big)\Big] \\ \nonumber
&-& {1 \over 3}\Big[ \Big( \delta_{ac}(\tilde \phi^{-1} \partial_v \tilde \phi) + V_{va} (\tilde \phi^{-1} \partial_r \tilde \phi) V_{rc} \Big) [ V_{sa} V_{ud} \partial_u \Gamma^{(1)}_a + \delta_{ad} \partial_s \Gamma^{(1)}_a - {Q^2_d \over Q^2_a} V_{sd} V_{ta} \partial_t \Gamma^{(1)}_a ] + \\ \nonumber
&&+ \Big( V_{sa} V_{ud} (\tilde \phi^{-1} \partial_u \tilde \phi) + \delta_{ad} (\tilde \phi^{-1} \partial_s \tilde \phi) \Big) [ \delta_{ac} \partial_v \Gamma^{(1)}_a + V_{va} V_{rc} \partial_r \Gamma^{(1)}_a - {Q^2_c \over Q^2_a} V_{vc} V_{ta} \partial_t \Gamma^{(1)}_a]+\\ \nonumber
&& - {Q^2_b \over Q^2_a} V_{ta} (\tilde \phi^{-1} \partial_t \tilde \phi) [\delta_{cb} (V_{ud} V_{sa} \partial_u \Gamma^{(1)}_a + \delta_{da} \partial_s \Gamma^{(1)}_a) + \delta_{db} (\delta_{ac} \partial_v \Gamma^{(1)}_a + V_{ud} V_{va} \partial_r \Gamma^{(1)}_a)]+ \\ \nonumber
&&- {Q^2_c Q^2_d\over Q^4_a} \, 2 \Big( (V_{sd} V_{vc}) V_{ta} V_{wa} (\tilde \phi^{-1} \partial_t \tilde \phi) \partial_w \Gamma^{(1)}_a \Big)\Big]+ \\ \nonumber
&-& \Big[ \delta_{ac} \partial_v \Gamma^{(1)}_a + V_{va} V_{rc} \partial_r \Gamma^{(1)}_a - {Q^2_c \over Q^2_a} V_{vc} V_{ta} \partial_t \Gamma^{(1)}_a \Big] \Big[ V_{sa} V_{ud}  \partial_u \Gamma^{(1)}_a + \delta_{ad} \partial_s \Gamma^{(1)}_a - {Q^2_d \over Q^2_a}  V_{ds} V_{wa} \partial_w \Gamma^{(1)}_d \Big] +\\ \nonumber
&-& {1 \over 6} \Big[ \Big[ \delta_{ac} \Big( \tilde \phi^{-1} \partial_v \tilde \phi \Big) + V_{va} V_{rc} (\tilde \phi^{-1} \partial_r \tilde \phi) - {Q^2_c \over Q^2_a} V_{vc} V_{ta} \Big( \tilde \phi^{-1} \partial_t \tilde \phi \Big) \Big] \\ \nonumber
&& \Big[ V_{ud} \partial_s V_{ua} + V_{ud} \partial_u V_{sa} + {Q^2_e \over Q^2_a} V_{wa} [ V_{ud} V_{se} (\partial_u V_{wd} - \partial_w V_{ue}) + \delta_{de} (\partial_s V_{we} - \partial_w V_{se})] \Big] +\\ \nonumber
\end{eqnarray*}

\begin{eqnarray*}
&&+ \Big[ V_{sa} V_{ud} \Big( \tilde \phi^{-1} \partial_u \tilde \phi \Big) + \delta_{ad} \Big( \tilde \phi^{-1} \partial_s \tilde \phi \Big) - {Q^2_d \over Q^2_a} V_{sd} V_{ta} \Big( \tilde \phi^{-1} \partial_t \tilde \phi \Big) \Big] \\ \nonumber
&& \Big[V_{rc} \partial_r V_{va} + V_{rc} \partial_v V_{ra} + {Q^2_e \over Q^2_a} V_{wa} [ \delta_{ce} (\partial_v V_{we} - \partial_w V_{ve}) + V_{rc} V_{ve}(\partial_r V_{we} - \partial_w V_{re}) ] \Big] \Big] \\ \nonumber
&-& {1 \over 2} \Big[ \Big[ \delta_{ca} \partial_v \Gamma^{(1)}_a + V_{ud} V_{rc} \partial_r \Gamma^{(1)}_a - {Q^2_c \over Q^2_a} V_{vc} V_{ta} \partial_t \Gamma^{(1)}_c \Big] \\ \nonumber
&&\Big[ V_{ud} \partial_s V_{ua} + V_{ud} \partial_u V_{sa} + {Q^2_e \over Q^2_a} V_{ud} V_{wa} [V_{se} (\partial_u V_{we} - \partial_w V_{ue}) + \delta_{de} (\partial_s V_{we} - \partial_w V_{se})] \Big] +\\ \nonumber
&&+ \Big[ V_{ud} V_{sa} \partial_u \Gamma^{(1)}_a + \delta_{da} \partial_s \Gamma^{(1)}_a - {Q^2_d \over Q^2_a} V_{sd} V_{ta} \partial_t \Gamma^{(1)}_d \Big] \\ \nonumber
&& \Big[ V_{rc} \partial_r V_{va} + V_{rc} \partial_v V_{ra} + {Q^2_e \over Q^2_a} V_{wa} [\delta_{ce} (\partial_v V_{we} - \partial_w V_{ve}) + V_{rc} V_{ve} (\partial_r V_{we} - \partial_w V_{re})] \Big]\Big] +\\ \nonumber
&-&{1 \over 4} V_{rc} V_{ud} \Big[ \partial_r V_{va} + \partial_v V_{ra} + {Q^2_e \over Q^2_a} V_{ta} [ V_{re} (\partial_v V_{te} - \partial_t V_{ve}) + V_{ve} (\partial_r V_{te} - \partial_t V_{re})] \Big] \\ \nonumber
&& \Big[ \partial_s V_{ua} + \partial_u V_{sa} + {Q^2_f \over Q^2_a} V_{wa} [V_{sf} (\partial_u V_{wf} - \partial_w V_{uf}) + V_{uf} (\partial_s V_{wf} - \partial_w V_{sf})]\Big]\Big\}
 \end{eqnarray*}

\begin{eqnarray*}
 {}^3R &=& {1 \over 2}\, \tilde \phi^{-2/3} \sum_a Q^2_a \sum_{cd} Q_c^{-2} Q_d^{-2} \Big( \delta_{(c)(d)} - {\bar{n}_{(c)} \bar{n}_{(d)} \over (1+n)^2} \Big) \sum_{ef} Q_e^{-2} Q_f^{-2} \Big( \delta_{(e)(f)} - {\bar{n}_{(e)} \bar{n}_{(f)} \over (1+n)^2} \Big) \\ \nonumber
& \Big\{& {2 \over 3} \Big[ V_{rc} \delta_{ae} \Big( \partial_r (\tilde \phi^{-1} \partial_u \tilde \phi) \delta_{af} V_{ud} - \partial_r (\tilde \phi^{-1} \partial_v  \tilde \phi) \delta_{ad} V_{vf} \Big) + \\ \nonumber
&&+ V_{se} \delta_{ac} \Big( \partial_s (\tilde \phi^{-1} \partial_s \tilde \phi) \delta_{ad} V_{vf} - \partial_s (\tilde \phi^{-1} \partial_u \tilde \phi) \delta_{af} V_{ud} \Big)\Big] + \\ \nonumber
&& + 2 \,V_{ud} V_{vf} \Big( V_{rc} \delta_{ae} (\delta_{af} \partial_r \partial_u \Gamma^{(1)}_a - \delta_{ad} \partial_r \partial_v \Gamma^{(1)}_a) + V_{se} \delta_{ac} (\delta_{ad} \partial_s \partial_v \Gamma^{(1)}_a - \delta_{af} \partial_s \partial_u \Gamma^{(1)}_a) \Big) + \\ \nonumber
&&+ 4 \Big[ \Big( {1 \over 3} (\tilde \phi^{-1} \partial_r \tilde \phi) + \partial_r \Gamma^{(1)}_a \Big) \delta_{ae} V_{rc} - \Big( {1 \over 3} (\tilde \phi^{-1} \partial_s \tilde \phi) + \partial_s \Gamma^{(1)}_a \Big) \delta_{ac} V_{se} \Big] \\ \nonumber
&&\Big[ \Big( {1 \over 3} (\tilde \phi^{-1} \partial_u \tilde \phi) + \partial_u \Gamma^{(1)}_a \Big) \delta_{fa} V_{ud} - \Big( {1 \over 3} (\tilde \phi^{-1} \partial_v \tilde \phi) + \partial_v \Gamma^{(1)}_a \Big) \delta_{ac} V_{vf} \Big] +\\ \nonumber
&&+ 2 V_{se} V_{vf} V_{rc} V_{vd} \Big[ \Big( {1 \over 3} \tilde \phi^{-1} \partial_r \tilde \phi + \partial_r \Gamma^{(1)}_a \Big) \Big( \partial_u (V_{sa}V_{va}) - \partial_v(V_{sa} V_{ua}) \Big) + \\ \nonumber
&&+\Big( {1 \over 3} \tilde \phi^{-1} \partial_s \tilde \phi + \partial_s \Gamma^{(1)}_a \Big) \Big( \partial_v(V_{ra}V_{ua} - \partial_u(V_{ra} V_{va}) \Big) + \\ \nonumber
&&+\Big( {1 \over 3} \tilde \phi^{-1} \partial_u \tilde \phi  + \partial_u \Gamma^{(1)}_a\Big) \Big( \partial_r (V_{sa} V_{va}) - \partial_s(V_{ra} V_{va}) \Big) + \\ \nonumber
&&+ \Big( {1 \over 3} \tilde \phi^{-1} \partial_v \tilde \phi + \partial_v \Gamma^{(1)}_a \Big) \Big( \partial_s(V_{ra} V_{ua}) - \partial_r(V_{va} V_{ua}) \Big)\Big] + \\ \nonumber
&&+ V_{se} V_{vf} V_{rc} V_{ud} \Big[ \partial_r \partial_u(V_{sa}V_{va}) - \partial_r \partial_v(V_{sa}V_{ua}) + \partial_s \partial_v (V_{ra}V_{ua}) - \partial_s \partial_u (V_{ra} V_{va}) \Big] + \\ \nonumber
\end{eqnarray*}

\bea
&-& {1 \over 9} \Big[ \Big( V_{vf} \delta_{ac}(\tilde \phi^{-1} \partial_v \tilde \phi) + V_{rc} \delta_{af}(\tilde \phi^{-1} \partial_r \tilde \phi) \Big) \Big( \delta_{ae} V_{ud} (\tilde \phi^{-1} \partial_u \tilde \phi) + \delta_{ad} V_{se} (\tilde \phi^{-1} \partial_s \tilde \phi) \Big) + \\ \nonumber
&&+ {Q^2_c Q^2_d \over Q_a^4} \delta_{cf} \delta_{ed} V_{ta} V_{wa} (\tilde \phi^{-1} \partial_t \tilde \phi) (\tilde \phi^{-1} \partial_w \tilde \phi) + \\ \nonumber
&&- {Q^2_b \over Q^2_a} V_{ta} (\tilde \phi^{-1} \partial_v \tilde \phi) \Big( \delta_{eb} V_{ub} (V_{ra} V_{vf} \tilde \phi^{-1} \partial_v \tilde \phi + \delta_{af} \tilde \phi^{-1} \partial_r \tilde \phi) + V_{rb} \delta_{bf} (\delta_{ae} \tilde \phi^{-1} \partial_u \tilde \phi + V_{se} V_{ua} \tilde \phi^{-1} \partial_s \tilde \phi)\Big)\Big] + \\ \nonumber
&-& {1 \over 3}\Big[ \Big( V_{vf} \delta_{ac}(\tilde \phi^{-1} \partial_v \tilde \phi) + \delta_{af} (\tilde \phi^{-1} \partial_r \tilde \phi) V_{rc} \Big) [ \delta_{ae} V_{ud} \partial_u \Gamma^{(1)}_a + \delta_{ad} V_{se} \partial_s \Gamma^{(1)}_a - {Q^2_d \over Q^2_a} \delta_{de} V_{ta} \partial_t \Gamma^{(1)}_a ] + \\ \nonumber
&&+ \Big( \delta_{ae} V_{ud} (\tilde \phi^{-1} \partial_u \tilde \phi) + \delta_{ad} V_{se} (\tilde \phi^{-1} \partial_s \tilde \phi) \Big) [ \delta_{ac} V_{vf} \partial_v \Gamma^{(1)}_a + \delta_{af} V_{rc} \partial_r \Gamma^{(1)}_a - {Q^2_c \over Q^2_a} \delta_{cf} V_{ta} \partial_t \Gamma^{(1)}_a]+\\ \nonumber
&&- {Q^2_c Q^2_d\over Q^4_a} \, 2 \Big( \delta_{cf} \delta_{ed} V_{ta} V_{wa} (\tilde \phi^{-1} \partial_t \tilde \phi) \partial_w \Gamma^{(1)}_a \Big)\Big]+ \\ \nonumber
&-& \Big[ V_{vf} \delta_{ac} \partial_v \Gamma^{(1)}_a + \delta_{af} V_{rc} \partial_r \Gamma^{(1)}_a - {Q^2_c \over Q^2_a} \delta_{cf} V_{ta} \partial_t \Gamma^{(1)}_a \Big] \Big[ \delta_{ea} V_{ud} \partial_u \Gamma^{(1)}_a + V_{se} \delta_{ad} \partial_s \Gamma^{(1)}_a - {Q^2_d \over Q^2_a}  \delta_{ed} V_{wa} \partial_w \Gamma^{(1)}_d \Big] +\\ \nonumber
&-& {1 \over 6} \Big[ \Big[ \delta_{ac} V_{vf} \Big( \tilde \phi^{-1} \partial_v \tilde \phi \Big) V_{ud} + \delta_{af} V_{rc} V_{ud} (\tilde \phi^{-1} \partial_r \tilde \phi) - {Q^2_c \over Q^2_a} \delta_{cf} V_{ud} V_{ta} \Big( \tilde \phi^{-1} \partial_t \tilde \phi \Big) \Big] \\ \nonumber
&& \Big[ V_{se} \partial_s V_{ua} + V_{se} \partial_v V_{sa} + {Q^2_h \over Q^2_a} V_{wa} [\delta_{eh} (\partial_u V_{wd} - \partial_w V_{uh}) + V_{se} V_{uh} (\partial_s V_{wh} - \partial_w V_{se})] \Big] +\\ \nonumber
&&+ \Big[ \delta_{ae} V_{rc} V_{ud} \Big( \tilde \phi^{-1} \partial_u \tilde \phi \Big) + V_{se} \delta_{ad} V_{rc} \Big( \tilde \phi^{-1} \partial_s \tilde \phi \Big) - {Q^2_d \over Q^2_a} \delta_{de} V_{rc} V_{ta} \Big( \tilde \phi^{-1} \partial_t \tilde \phi \Big) \Big] \\ \nonumber
&& \Big[ V_{vf} \partial_r V_{va} + V_{vf} \partial_v V_{ra} + {Q^2_h \over Q^2_a} V_{wa} [ V_{rh} V_{vf} (\partial_v V_{wh} - \partial_w V_{vh}) + \delta_{fh}(\partial_r V_{wh} - \partial_w V_{rh}) ] \Big] \Big] + \\ \nonumber
&-& {1 \over 2} \Big[ \Big[ \delta_{ca} V_{vf} V_{vd} \partial_v \Gamma^{(1)}_a + \delta_{da} V_{vf} V_{rc} \partial_r \Gamma^{(1)}_a - {Q^2_c \over Q^2_a} \delta_{cd} V_{vf} V_{ta} \partial_t \Gamma^{(1)}_c \Big]  \nonumber \\
&&\Big[ V_{se} \partial_s V_{ua} + V_{se} \partial_u V_{sa} + {Q^2_h \over Q^2_a} V_{wa} [\delta_{de} (\partial_u V_{wh} - \partial_w V_{uh}) + V_{se} V_{ud} (\partial_s V_{wh} - \partial_w V_{sh})] \Big] + \nonumber \\
&&+ \Big[V_{rc} V_{ud} (\delta_{ae} \partial_u \Gamma^{(1)}_a + V_{ua} V_{se} \partial_s \Gamma^{(1)}_a) - {Q^2_d \over Q^2_a} \delta_{de} V_{rc} V_{ta} \partial_t \Gamma^{(1)}_d \Big]  \nonumber \\
&& \Big[ V_{vf} \partial_r V_{va} + V_{vf} \partial_v V_{ra} + {Q^2_h \over Q^2_a} V_{wa} [V_{rh} V_{vf} (\partial_v V_{wh} - \partial_w V_{vh}) + \delta_{fh} (\partial_r V_{wh} - \partial_w V_{rh})] \Big]\Big] + \nonumber \\
&-&{1 \over 4} V_{rc} V_{ud} V_{se} V_{vf} \Big[ \partial_r V_{va} + \partial_v V_{ra} + {Q^2_h \over Q^2_a} V_{ta} [ V_{rh} (\partial_v V_{th} - \partial_t V_{vh}) + V_{ve} (\partial_r V_{th} - \partial_t V_{rh})] \Big] \nonumber \\
&& \Big[ \partial_s V_{ua} + \partial_u V_{sa} + {Q^2_i \over Q^2_a} V_{wa} [V_{si} (\partial_u V_{wi} - \partial_w V_{ui}) + V_{ui} (\partial_s V_{wi} - \partial_w V_{si})]\Big]\Big\}.\nonumber \\
{}&&
 \label{a1}
 \eea

\vfill\eject

\end{document}